\pdfoutput=1
\documentclass[aps,prd,amsmath,floats,floatfix, twocolumn,
superscriptaddress,nofootinbib,showpacs,longbibliography]{revtex4-1}

\usepackage[T1]{fontenc}
\usepackage[utf8]{inputenc}
\usepackage{lmodern}
\usepackage{mathtools}
\usepackage{times}
\usepackage[varg]{txfonts}
\usepackage{verbatim}
\usepackage{multirow}
\usepackage{longtable}
\usepackage{booktabs}

\usepackage{fontawesome5}
\usepackage{array}
\usepackage[dvipsnames, usenames, table, svgnames]{xcolor}
\definecolor{linkcolor}{rgb}{0.0,0.3,0.5}
\usepackage[hypertexnames=false, unicode, colorlinks=true, linkcolor=linkcolor,
citecolor=linkcolor, filecolor=linkcolor,urlcolor=linkcolor,
pdfusetitle]{hyperref}

\usepackage{float} 

\makeatletter
\g@addto@macro{\endtabular}{\rowfont{}}
\makeatother

\newcommand{\rowfonttype}{}
\newcommand{\rowfont}[1]{
\gdef\rowfonttype{#1}#1\ignorespaces%
}
\makeatother

\usepackage[all]{hypcap}
\usepackage{graphicx}
\usepackage{geometry}
\usepackage{xspace}
\usepackage{amssymb}
\usepackage{amsmath} 
\usepackage[normalem]{ulem} 
\usepackage{bm} 
\usepackage{enumitem}

\usepackage{microtype}
\usepackage[english]{babel}
\usepackage{blindtext}
\usepackage{orcidlink}
\usepackage{setspace}
\graphicspath{%
  {figs/}%
}

\DeclareMathAlphabet{\mathpzc}{OT1}{pzc}{m}{it}

\newcommand{\h}{\mathpzc{h}}
\newcommand{\hlm}{\mathpzc{h}_{\ell m}}

\newcommand{\dd}{\mathrm{d}}

\begin{document}

\title{A comparison between best-fit eccentricity definitions and the standardized definition of eccentricity}

\newcommand{\nico}[1]{{\leavevmode\color{red}\textbf{(NC: #1)}}}
\newcommand{\arif}[1]{{\leavevmode\color{orange}{(AS: #1)}}}
\newcommand{\jeongcho}[1]{{\leavevmode\color{blue}{(JK: #1)}}}
\newcommand{\hml}[1]{{\leavevmode\color{magenta}{(HML: #1)}}}

\newcommand{\code}[1]{\texttt{#1}}
\newcommand{\emphasis}[1]{\textit{#1}}

\newcommand{\eBF}{e^{\text{best-fit}}}
\newcommand{\egwBF}{e_{\text{gw}}^{\text{best-fit}}}
\newcommand{\egwNR}{e_{\text{gw}}^{\text{NR}}}
\newcommand{\fBF}{f^{\text{best-fit}}}
\newcommand{\hBF}{\h^{\text{best-fit}}}
\newcommand{\hmodel}{\h_{\text{model}}}
\newcommand{\hNR}{\h_{\text{NR}}}
\newcommand{\omegaP}{\omega_{22}^{\text{p}}}
\newcommand{\omegaA}{\omega_{22}^{\text{a}}}
\newcommand{\ie}{\emph{i.e.}\xspace}
\newcommand{\eg}{\emph{e.g.}\xspace}
\newcommand{\mode}{$\left(2, |2|\right)$\xspace}
\newcommand{\teob}{\code{TEOBResumS-Dal{\'i}}\xspace}
\newcommand{\seob}{\code{SEOBNRE}\xspace}
\newcommand{\mm}{\mathcal{M}\xspace}
\newcommand{\mismatch}{\mathcal{M}\xspace}
\newcommand{\match}{1-\mm \xspace}
\newcommand{\hL}{\mathcal{E}_{\h_{22}}\xspace}
\newcommand{\omegaL}{\mathcal{E}_{\omega_{22}}\xspace}
\newcommand{\egw}{e_{\text{gw}}}
\newcommand{\etwotwo}{e_{\omega_{22}}}
\newcommand{\degw}{\delta e_{\text{gw}}}
\newcommand{\degwd}{\Delta e_{\text{gw}}}
\newcommand{\hres}{\|\delta h_{22}\|}
\newcommand{\stone}{\textsc{Rosetta}\xspace}
\newcommand{\Python}{\texttt{Python}\xspace}
\newcommand{\gwecc}{\texttt{gw\_eccentricity}\xspace}
\newcommand{\resamp}{\texttt{ResidualAmplitude}\xspace}

\newcommand{\bmt}{\bm{\theta}}

\newcommand{\github}[1]{%
   \href{#1}{\faGithubSquare}%
}

\newcommand{\SNU}{\affiliation{Department of Physics and Astronomy,
    Seoul National University, Seoul 08826, Republic of Korea}}
\newcommand{\seoul}{Center for the Gravitational-Wave Universe, Astronomy Program Department of Physics and Astronomy, Seoul National University, Seoul 08826, Korea}
\newcommand{\NIMS}{\affiliation{Division of Fundamental Researches on Public Agendas,
National Institute for Mathematical Sciences, 34047, Daejeon, Republic of Korea}}
\newcommand{\VSM}{\affiliation{Department of Physics, Vivekananda Satavarshiki
    Mahavidyalaya (affiliated to Vidyasagar University), Manikpara 721513, West
    Bengal, India}}

\author{Nicolas Chartier}
\SNU

\author{Md Arif Shaikh\,\orcidlink{0000-0003-0826-6164}}
\email{arifshaikh.astro@gmail.com}
\SNU
\VSM

\author{Hyung Mok Lee}
\SNU

\author{JeongCho Kim}
\NIMS

\hypersetup{pdfauthor={Chartier et al.}}

\date{\today}

\begin{abstract}
In the absence of a unique, gauge-independent definition of eccentricity in General Relativity, there have been efforts to standardize the definition for Gravitational-Wave astronomy. Recently, Shaikh et al. proposed a model-independent measurement of eccentricity $\egw$ from the phase evolution of the dominant mode.
Many works use loss functions (LFs) to assign eccentricity to a reference waveform, for instance by fitting a Post-Newtonian expression to assign eccentricity to Numerical Relativity (NR) simulations.
Therefore, we ask whether minimizing common LFs on gauge-dependent model parameters, such as the mismatch $\mm$ or the $L_2$-norm of the dominant mode $\h_{22}$ residuals, for non-precessing binaries, ensures a sufficient $\egw$ agreement.
We use $10$ eccentric NR simulations and the eccentric waveform \teob as the parametric model to fit on eccentricity  $e_0$ and reference frequency $f_0$. We first show that a minimized mismatch, the  $\mm \sim 10^{-3}- 10^{-2}$ results in better $\egw$ fractional differences ($\sim 1\%$) than with the minimized $\h_{22}$ residuals. Nonetheless, for small eccentricity NR simulations $(\egw \lesssim 10^{-2}$), the mismatch can favor quasi-circular ($e_0=0$) best-fit models. Thus, with sufficiently long NR simulations, we can include $\egw$ in the LF. We explain why solely fitting with $\egw$ constitutes a degenerate problem. 
To circumvent these limitations, we propose to minimize a convex sum of $\mm$ and the $\egw$ difference to both assign non-zero eccentric values to NR strains and to control the mismatch threshold.
\end{abstract}
\maketitle

\section{Introduction}
\label{sec:introduction}

Gravitational Wave (GW) observations have opened up a new window to the
universe. The ground-based network of GW observatories
LIGO-Virgo~\cite{TheLIGOScientific:2014jea, TheVirgo:2014hva} has detected
about 90 compact binary coalescences (CBCs) in the first three observing
runs~\cite{KAGRA:2021vkt} that includes detection of binary black hole (BBH),
black hole-neutron star (BHNS)~\cite{LIGOScientific:2021qlt} and binary neutron
star (BNS)~\cite{TheLIGOScientific:2017qsa, Abbott:2020uma} systems. One of the
key questions that we hope to address using these detections is how these
binaries form in nature, and the eccentricity of the binary orbit is one of the indicators that could play an important role in this regard.

Binaries lose orbital eccentricity due to the loss of energy and angular
momentum via GW radiation~\cite{Peters:1963ux, Peters:1964zz}.  It is expected
that for binaries formed via isolated evolution (e.g., in the galactic fields
~\cite{Mapelli:2021for}), the orbit of the binary will become circularized by
the time the GW emitted from such CBCs enters the sensitive frequency band of
the ground-based GW detectors. Indeed, so far all the detection and analyses of
the CBC events by the LIGO-Virgo-KAGRA (LVK)~\cite{TheLIGOScientific:2014jea,
TheVirgo:2014hva, KAGRA:2020tym} collaboration has been performed assuming the
binary to be in a quasicircular orbit~\cite{LIGOScientific:2021djp}). However,
recent studies indicate that the GW from binaries formed in a dense environment
via dynamical formation~\cite{Samsing:2020tda, tagawa2021eccentric,
Kozai:1962zz, Lidov:1962wjn, Naoz:2016tri, Antonini:2017ash, Randall:2017jop,Zevin:2018kzq,
Yu:2020iqj, Bartos:2023lfu, Bhaumik:2024cec} (e.g., in globular clusters, galactic nuclei, field
triples etc.) could retain non-negligible eccentricity when entering the
sensitive band of the ground-based GW detectors. Thus detection of
non-negligible eccentricity could indicate the dynamical formation of the
binary. For instance, debates as to the characteristics and eccentricity of \texttt{GW190521}, along with other merger events, are still ongoing \cite{Romero-Shaw:2020thy, Gayathri:2020coq, Romero-Shaw:2021ual, Siegel:2023lxl, Gupte:2024jfe}.

Consequently, the GW community has undertaken significant efforts to develop tools and
frameworks for detecting and analyzing GW data from such binaries on eccentric
orbits. Several groups have been actively developing gravitational waveform
models for binaries in eccentric orbit using semi-analytic formalism like
Effective-One-Body (EOB) and Post-Newtonian approximation as well as employing
fully Numerical Relativity (NR) simulations~\cite{Warburton:2011fk,
Osburn:2015duj, Cao:2017ndf, Liu:2019jpg, Ramos-Buades:2021adz, Nagar:2021gss,
Islam:2021mha, Liu:2021pkr, Memmesheimer:2004cv, Huerta:2014eca, Tanay:2016zog,
Cho:2021oai, Moore:2018kvz, Moore:2019xkm, VanDeMeent:2018cgn, Chua:2020stf,
Hughes:2021exa, Katz:2021yft, Lynch:2021ogr, Klein:2021jtd,Wang:2024jro, Shi:2024age, Phukon:2024amh,Sridhar:2024zms, Gamboa:2024imd, Gamboa:2024hli, Morras:2025nlp, Planas:2025feq}. ,  Several works demonstrated the need to include eccentricity by
showing potential bias in estimated parameters and spurious violation of
General Relativity due to neglecting eccentricity~\cite{Favata:2021vhw,OShea:2021faf,Saini:2022igm, Saini:2023rto,Favata:2013rwa,Shaikh:2024wyn, Das:2024zib, Kumar:2025nwb}. The inclusion of other physical effects of eccentric binaries, such as the argument of periapsis \cite[\eg][]{Clarke:2022fma}, is also being discussed.

Modelling waveforms from compact binaries in eccentric orbits has one challenging aspect. Unlike Newtonian orbits, the binary orbit in a strong gravity regime can not be described by a closed ellipse -- the orbit goes through pericenter precession. In addition, due to GW radiation, the size of
the orbit and the eccentricity reduce towards merger time~\cite{Zevin:2018kzq}. There is no
unique definition of eccentricity in General Relativity and, therefore, to capture features of a binary in an eccentric orbit, different waveform models adopt internal definitions as per the convenience of the
model framework. This leads to an unwanted ambiguity regarding the eccentricity parameter of waveform models~\cite{Knee:2022hth}. Such
ambiguity could be reflected in the posteriors of parameter inference from GW strain data that include eccentricity, and would impede our understanding of astrophysical populations among research groups choosing different eccentric waveform models.

To avoid such ambiguity, there has been efforts to standardize the definition of eccentricity based on a
gauge-independent quantity like the GW modes at
infinity~\cite{Ramos-Buades:2019uvh, Islam:2021mha,
Bonino:2022hkj, Ramos-Buades:2022lgf, Shaikh:2023ypz}. Since such a definition uses only the information in the GW modes, it removes the model dependency,
\ie it is independent of the way the GW modes are generated. This definition makes use of the oscillatory nature of the frequency and amplitude evolution of
the eccentric GWs--it determines the local maxima and minima from
waveform quantities like the frequency or the amplitude. Consequently,
the computation requires a minimum
(typically $\gtrsim 5$) number of orbits  in the waveform to reliably estimate eccentricity. Thus far, the definition has been examined primarily for waveforms in which the $(2,2)$ multipole is dominant.
See Ref.~\cite{Shaikh:2023ypz} for more details on strategies to implement a
standardized definition of eccentricity, which we refer to as $\egw$. Recently, \cite{Patterson:2024vbo} proposed an eccentricity identification, also restricted to the $(2,2)$-mode, from an harmonic decomposition of the waveform and \cite{Boschini:2024scu} developed a gauge independent estimation method from catastrophe theory. The authors of \cite{Islam:2025oiv} propose a publicly available framework for smooth, monotonic eccentricity estimates invoking universal modulation features of non-precessing binaries \cite{Islam:2024rhm, Islam:2024bza}.

Although advances in NR formalism and large computational resources have enabled NR groups to simulate increasingly long compact binary coalescences (CBCs), including eccentric ones, many NR simulations lack a sufficient number of orbits to reliably apply the standardized definition of eccentricity. For such shorter NR simulations, other definitions of eccentricity may be used to assign an eccentricity value to these
simulations. One commonly used definition of eccentricity involves 
the minimization of a Loss Function (LF) by adjusting the parameters of a chosen waveform model to best match the given NR waveform. The \textit{mismatch} $\mathcal{M}$ \cite{Flanagan:1997kp}\footnote{or equivalently its counterpart to maximize, the \textit{overlap} or \textit{faithfulness} \eg in \cite{Hinder:2017sxy}}, is commonly used in such procedure since this scalar quantity defines the optimal matched filter in the GW detection problem. The mismatch can either play the role of the LF to minimize  \cite[\eg][]{Joshi:2022ocr} or can be used as a validation metric after minimization of another LF. For instance, the authors of \cite{Habib:2019cui} computed their LFs from the evolution of the dimensionless quantity $M\omega$, with $M$ the BBH total mass and $\omega$ the mean orbital frequency, and then calculated the mismatch to quantify the accuracy of the characterization of the eccentric NR simulation at hand.

Moreover, the mismatch is also routinely used as the LF to validate eccentric waveform models against NR simulations~\cite[\eg][]{Hinder:2017sxy,Liu:2019jpg,  Nagar:2021gss,Bonino:2022hkj}. 
Other applications of the mismatch include mapping the eccentricity definitions between distinct waveform models, as \cite{Knee:2022hth} extensively studied for the models \code{TEOBResumS-Dali}~\cite{Chiaramello:2020ehz, Nagar:2021xnh} and \code{SEOBNRE} \citep{seobnre}, or developing fast surrogate models \cite{Blackman:2015pia,Blackman:2017dfb,Blackman:2017pcm, Islam:2021mha}. For instance, \cite{Yun:2021jnh} fitted a reduced-order (RO) surrogate for \seob, covering eccentricities up to $\sim 0.25$ to improve GW inference computational capabilities.  We stress that these works used the mismatch as a validation tool for the fitted surrogates, not to fit the surrogate models beforehand. 

Alternatively, one can choose the $L_2$-norm, the distance between vectors of strain data in time domain as the LF to compare a model (waveform, surrogate model to train) from a set of NR simulations \cite{Blackman:2017dfb, Islam:2021mha,Islam:2022laz, Islam:2024tcs, Rink:2024swg}. We will thus also include this $L_2$-norm in our analysis.

However, a minimized mismatch, by definition,
makes sure that the waveform modes in frequency
domain are best correlated, and does not necessarily
guarantee an optimal agreement of the eccentricity
evolution independently computed from the NR and
the model waveform as per the standardized definition. 
This limitation motivates us to ask whether these two definitions are equivalent in practice, \ie does the eccentricity evolution $\egw$ 
independently computed from an NR waveform match the measured $\egw$ of a model waveform parametrized so that it minimizes a LF? We address this question by taking a set of NR waveforms and computing the difference in the eccentricity between these two approaches. We consider NR waveforms that are long enough so that they contain enough number of orbits for the standardized
definition to be applicable. Studying the case of \teob for which eccentricity is defined with two parameters -- the initial eccentricity $e_0 \in [0,1.]$ and reference frequency $f_0$ -- with numerical search strategies similar \cite{Knee:2022hth}, we find that there does {\itshape not} exist a simple one-to-one correspondence between the best match waveforms and the
eccentricity evolution. A lower mismatch is not equivalent to a better $\egw$ agreement, and conversely we find that fitting for eccentricity agreement on $\egw$ results in degenerate solutions and is thus not sufficient to characterize eccentric NR simulations. 
In particular, we will interpret the previous points regarding the limitations of a model allowing to define eccentricity through $e_0$ and $f_0$ only, without the mean anomaly $l_0$. We will also discuss the very recent findings of \cite{Bonino:2024xrv} who used the standardized $\egw$ definition to map eccentricity between NR simulations and waveform models. The authors of this work, proposed to determine new pairs of initial conditions based on the time difference of the first apastra for \teob or similar EOB waveforms.

We organize the study as follows. In Section \ref{sec:methods}, we introduce some useful reminders about the mathematical formalism for waveforms and about eccentricity in GW astronomy. The next section deals with the characterization of NR eccentric simulations using the mismatch and other LFs with a given waveform approximant. We incorporate the measurement of eccentricity as developed in \cite{Shaikh:2023ypz} in this process, and emphasize the limitations of the mismatch to define eccentric waveforms and then propose an alternative, fitting procedure, assuming we can only modify $(e_0,f_0)$. Finally, we propose a new, simple LF to circumvent aforementioned limitations and we discuss the implications of our findings when it comes to assessing the agreement between eccentric approximants when relying solely on mismatch.

\section{Theoretical Background}
\label{sec:methods}

\subsection{Gravitational waveform and mismatch}
\label{subsec:formalism}
The plus ($h_+$) and cross ($h_{\times}$) polarizations of GWs can be combined
to define a complex strain $\h \equiv h_{+} -i h_{\times}$. The complex strain
$\h$ can be decomposed into a sum of spin-weighted spherical harmonics modes
$\hlm$ along any direction $(\iota, \varphi_0)$ in the binary's source frame as
\begin{equation}\label{eq:complex_polarizations} \h(t, \iota, \varphi_0) =
\sum_{\ell=2}^{\infty}\sum_{m=-\ell}^{\ell} \hlm(t) \prescript{}{-2}{Y}_{\ell
m}(\iota, \varphi_0)
\end{equation} where $\iota$ and $\varphi_0$ are the polar and azimuthal
angles, respectively, on the sky in the source frame and $_{-2}{Y}_{\ell m}$ is
the spin = -2 weighted spherical harmonics.

The mismatch $\mm$ between two given GW strains $\h_1$ and $\h_2$ is the bounded scalar
defined as \citep{CutlerFlanagan1994}
\begin{equation}
  \label{eq:mismatch}
  \mm = 1 - \arg \max_{t_c, \phi_c}\frac{\langle\h_1, \h_2\rangle}{\sqrt{\langle\h_1, \h_1\rangle\langle\h_2, \h_2\rangle}},
\end{equation}
where $t_c$ and $\phi_c$ are the coalescence time and phase, respectively, and
the inner product $\langle \h_1, \h_2\rangle$ is
\begin{equation}
  \label{eq:overlap}
  \langle \h_1, \h_2\rangle = 4 \mathcal{R} \int_{f_{\text{low}}}^{f_{\text{high}}} \frac{\tilde{\h_1}^\star(f) \tilde{\h}_2(f)}{S_n(f)}\dd f.
\end{equation}
Here, $f$ is the frequency, $\tilde{\h}_{1,2}$ are the Fourier transforms of
the corresponding time domain waveforms, $\star$ represents the complex
conjugate and $\mathcal{R}$ represents the real part. Further, $f_{\text{low}}$ and
$f_{\text{high}}$ denote the lower and higher frequency cutoff of the detector
band and $S_{n}(f)$ denote the one-sided power spectral density (PSD) of the
detector noise. We recognize in equation \eqref{eq:overlap} the inner product of square-integrable functions, weighted by $S_n$, and in the second term of equation \eqref{eq:mismatch} the correlation coefficient in $[-1,1]$ in Fourier space.

\subsection{Eccentricity definitions}
\label{subsec:eccentricity}


In the Newtonian regime, the orbital eccentricity can be uniquely defined using the orbital separations at apocenter ($r^{\text{a}}$) and pericenter ($r^{\text{p}}$)
\begin{equation}
  \label{eq:kepler}
  e_{\text{Newt}} = \frac{r^{\text{a}} - r^{\text{p}}}{r^{\text{a}} + r^{\text{p}}}\,,
\end{equation}
In the GR regime, the binary orbit is no longer a closed ellipse. Due to the GW radiation, the orbit loses energy and angular momentum, causing the orbital size to shrink as the system inspirals. In addition, GR effect induces pericenter precession. In the quasi-Keplerian formalism, these effects are captured by extending the Keplerian equation of motion through the introduction of different eccentricity parameters --- the radial ($e_r$), temporal ($e_t$) and the angular ($e_\phi$) eccentricities. These quantities are usually defined in terms of the conserved energy and the angular momentum which are \textit{gauge dependent}. In the EOB formalism, the initial conditions for the dynamics are prescribed in terms of an eccentricity parameter that is defined using the quasi-Keplerian formalism, making it a gauge dependent definition. Similarly, in the self-force calculations of extreme mass ratio inspirals, we characterize eccentricity using the turning points of geodesics which constitutes again a gauge dependent choice since it depends on underlying coordinate systems used to describe the background spacetime metric. Finally, for the NR simulations, we fit the compact object trajectories to analytical PN or Newtonian expressions to define eccentricity. Since the trajectories are dependent on the gauge used in the simulation, the subsequent eccentricity characterization is also gauge dependent.
\subsubsection{Standardized eccentricity definition}
\label{Standardized_eccentricity_definition}

To avoid the gauge ambiguity and model dependence, a definition of eccentricity
based solely on the gravitational waveform has been proposed ~\cite{Ramos-Buades:2021adz, Islam:2021mha, Ramos-Buades:2019uvh,
Bonino:2022hkj}. This eccentricity $\etwotwo$ is computed using the
interpolants through frequency of the (2, 2) mode at the apocenter ($\omegaA(t)$)
and pericenter ($\omegaP(t)$)
\begin{equation}
  \label{eq:e22}
  e_{\omega_{22}}(t) = \left(\frac{\sqrt{\omegaP(t)} -  \sqrt{\omegaA(t)}}{\sqrt{\omegaP(t)} +  \sqrt{\omegaA(t)}}\right)\,,
\end{equation}
where frequency $\omega_{22}$ is obtained from $\h_{22}$ using the following decomposition into amplitude $A_{22}$ and phase $\Phi_{22}$
\begin{equation}
  \label{eq:22mode}
  \begin{aligned}
    \h_{22} &= A_{22} e^{-i \Phi_{22}}\\
    \omega_{22} &=\frac{\dd\Phi_{22}}{\dd t}.
  \end{aligned}
\end{equation}



However, $e_{\omega_{22}}$  does not have the correct Newtonian order limit.
Therefore, to ensure the correct Newtonian limit, \cite{Shaikh:2023ypz} proposed a new waveform based definition of eccentricity $\egw$.

\begin{align}\label{eq:egw}
    \egw &= \cos(\Psi_{22}) - \sqrt{3}\sin(\Psi_{22}) \\
   \Psi_{22} &\equiv \frac{1}{3}\arctan\left( \frac{1 - e_{\omega_{22}}^2}{2e_{\omega_{22}}}\right)\,.
\end{align}

Because $\egw$ uses interpolants through frequency at the pericenters and
apocenters, this definition requires the waveform to contain a minimum number
of cycles ($\sim 5$) to build the interpolants robustly. Therefore, for shorter
waveforms, we might not be able to reliably compute $\egw$ and have to resort
to other methods. In many NR simulations, this could be the case where the
simulations have only a few final orbits of the compact object binary.



\subsubsection{Eccentricity of the best fitted waveform}
\label{Eccentricity_of_the_best_fitted_waveform}

One alternative way of assigning an eccentricity to an NR waveform is to find the initial data ($\fBF, \eBF$), where $\eBF$ is the initial eccentricity at the
initial frequency $\fBF$ of the best-fit waveform,

\begin{equation}
  \label{eq:eModel}
  (\fBF, \eBF) = \arg \min_{(f_0, e_0)} \mathcal{E}\left[\hNR, \hmodel(f_0,e_0)\right],
\end{equation}
where, the Loss Function (LF) $\mathcal{E}$ is a measure of disagreement between the
NR and the model generated waveform $\hmodel$ at $(f_0, e_0)$. For example, one usually
use the mismatch $\mm$ (defined in equation \eqref{eq:mismatch}) as a good representation of $\mathcal{E}$. We discuss possible choices for LFs in Section \ref{sec:loss_functions}.

This method, unlike equation \eqref{eq:egw}, uses an waveform model and therefore is
model dependent. However, it does not depend on the number of orbits contained
in the waveform and, therefore, can be applied to any waveform.

In the following sections, we test the robustness of the definition $\eBF$
in equation \eqref{eq:eModel} for assigning eccentricity to NR simulations when
compared to the $\egw$ computed from the NR waveform. We accomplish this by
studying a set of longer simulations where both of these methods are
applicable.

\section{Methodology and setup}
\label{Methodology_and_setup}

\subsection{NR simulations and model waveform}
\label{NR_and_model_waveform}
For this study, we use $10$ eccentric simulations, as listed in Table \ref{table:SXSdata}, from the Simulating eXtreme
Spacetimes (SXS) catalog \citep{Boyle:2019kee,
SXSCatalog, SXSWebsite} and run by \cite{Islam:2021mha}. We sort them in this table by $\egw$ at $18$Hz from the measurements of Fig.~\ref{fig:sxs_egw}.
We use the implementation in the \Python library
\gwecc~\footnote{\url{https://github.com/vijayvarma392/gw\_eccentricity}} from \cite{Shaikh:2023ypz} to compute $\egw$ in this work. We explain the chosen method later in Section \ref{subsec:measure_egw}.

All the NR simulations in Table.~\ref{table:SXSdata} have mass ratio $q \equiv
m_2/m_1 = 1 $ ($m_2 \leq m_1$). We convert the strain from geometric units to physical units by fixing the binary mass to $M = m_1 + m_2 = 60
M_{\odot}$ and its luminosity distance to $D_L = 400$ Mpc similar to the parameters of \texttt{GW150914}. 

To assign $\eBF$ to these simulations, we use the publicly available
eccentric waveform model \teob \footnote{\url{https://bitbucket.org/teobresums/teobresums/}. 
We have used the git commit \texttt{0f19532a9a1012b556bd4c72da10a1317d0e8751}}. To compute $\egw$ and the LF $\mathcal{E}$, we
will make use of the dominant (2, 2) mode only and output physical units strain from \teob, In addition, since \teob allows to define the reference frequency as the waveform frequency at apastron, periastron or the mean of the two, we choose the latter definition as in \cite{Knee:2022hth}.
The inclusion of higher modes has notably been investigated by \eg \cite{Habib:2019cui,Bonino:2022hkj}, and is beyond the scope of our study. It should also be kept in mind that the definition of the eccentric initial conditions are sensitive to changes in input parameters $e_0, f_0 \sim \mathcal{O}(10^{-6})$ \cite{Nagar:2021xnh}.


\newcolumntype{C}[1]{>{\centering\arraybackslash}p{#1}}
\begin{table}[h!]
   \caption{Basic information about our set of equal mass, eccentric NR simulations from SXS. We analyze the equal mass binaries with physical units corresponding to total mass $M = 60 M_{\odot}$ and  luminosity distance $D_L = 400$Mpc.}\label{table:SXSdata}
\begin{longtable}[c]{|C{2.2cm} |C{2.2cm} |C{2.2cm}|}
     \toprule
     Simulation ID  & $\lfloor N_{\text{orbit}} \rfloor$ & $\egw (18\mathrm{Hz})$ \\
      \toprule
          \texttt{SXS:BBH:2294}  & $27$ & 0.0028  \\
       \midrule
       \texttt{SXS:BBH:2267}  & $26$ & 0.011\\
      \midrule
     \texttt{SXS:BBH:2270} & $26$ & 0.027\\
      \midrule
     \texttt{SXS:BBH:2275}  & $26$ & 0.059 \\
       \midrule
    \texttt{SXS:BBH:2280}  & $26$ &  0.083\\
      \midrule
    \texttt{SXS:BBH:2285}  & $24$ & 0.103 \\
       \midrule
    \texttt{SXS:BBH:2290} & $21$  & 0.139\\
       \midrule
    \texttt{SXS:BBH:2311}  & $27$ &  0.168\\
      \midrule
    \texttt{SXS:BBH:2300}  & $24$ & 0.168 \\
       \midrule
    \texttt{SXS:BBH:2305}  & $25$ & 0.198 \\
       \bottomrule
\end{longtable}
\end{table}
\setcounter{table}{1}

\subsection{Measuring $\egw$}\label{subsec:measure_egw}
\begin{figure*}
\includegraphics[width=0.49\textwidth]{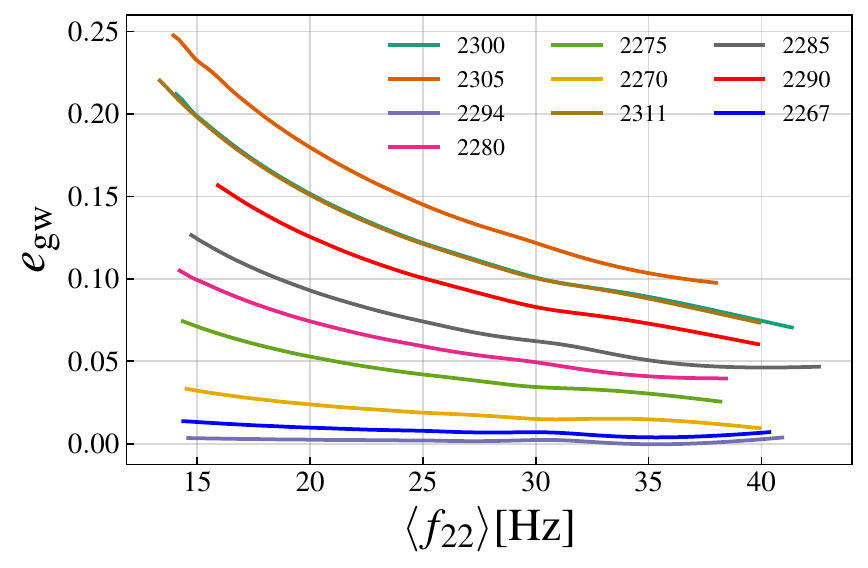}
\includegraphics[width=0.49\textwidth]{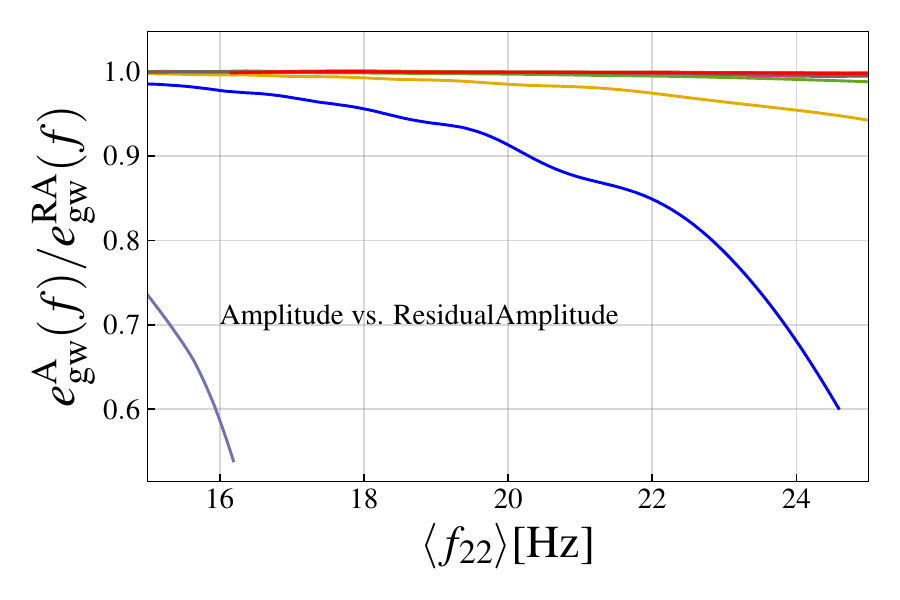}
    \caption{\underline{Left:} Frequency evolution of eccentricity $\egw$ in the simulations from
    Table.~\ref{table:SXSdata}.
    \underline{Right:} $\egw$ ratio between the \resamp and \texttt{Amplitude} method from \gwecc \cite{Shaikh:2023ypz}. The former is more robust to the small eccentric NR strains in our set, justifying our preference. We resort to physical units by choosing a total binary mass $M=60M_{\odot}$ and luminosity distance $D_L =
    400$Mpc.}
  \label{fig:sxs_egw}
\end{figure*}

For $\egw$ measurements, we consider the signals up to $2$ orbits before merger.
These measurements rely, as currently implemented, on interpolants through $\omega_{22}$ at apastron and periastron in equation \eqref{eq:e22}. The authors of Ref. \cite{Shaikh:2023ypz} propose different methods to locate apatron and periastron times, and for the sake of simplicity we will present only our choice for this work.
We could not measure $\egw$ from the straightforward \texttt{Amplitude} method, which locates extrema directly in $A_{22}(t)$, for small eccentricity cases. Instead, chose the \resamp method instead that is more robust for small eccentricities at the cost of using a quasi-circular counterpart\footnote{We found the computational cost to be lighter than with the \texttt{AmplitudeFits} or \texttt{FrequencyFits} methods also proposed in \gwecc.} to locate the extrema of $A_{22}(t) - A_{22}^{\mathrm{circ}}(t)$. The quasi-circular counterpart of \teob will simply be the same waveform computed with $e_0=0$, while we will use the default \texttt{IMRPhenomT} approximation to measure $\egw$ of our SXS data. Figure \ref{fig:sxs_egw} shows the $\egw$ measurements in our SXS set and emphasizes our motivation to prefer the \resamp, in accordance to the conclusions of the original work about small eccentricities. Indeed, with the \texttt{Amplitude} method, we cannot measure $\egw$ in a frequency range as large as with \resamp for \texttt{SXS:BBH:2294}, \texttt{SXS:BBH:2267} and \texttt{SXS:BBH:2270}, and the agreement ratio with \resamp is the lowest in these cases (right panel of Figure \ref{fig:sxs_egw}).

\subsection{Methodology}\label{subsec:methodology}
\subsubsection{Comparison study strategy}
\label{Comparison_study_strategy}
To test the closeness of $\eBF$, assigned to an NR simulation, when compared to
$\egwNR$ measured directly from the NR waveform, we use the following strategy:
\begin{enumerate}
\item We compute $\egwNR$ directly from the given NR waveform.
\item Using a waveform model and a LF, we find the best-fit waveform $\hBF$
  by finding the initial frequency $\fBF$ and eccentricity $\eBF$ by minimizing
  the LF using equation \eqref{eq:eModel}. In this work, we use \teob, as mentioned in
  Section \ref{NR_and_model_waveform}, for generating the model
  waveform $\hmodel$ to find $\hBF$. In Section \ref{sec:loss_functions}, we
  discuss different LFs that can be used to do such a study.
\item Next, using equation \eqref{eq:egw}, we compute $\egwBF$ of the best matched
  waveform $\hBF$ obtained in the previous step.
\item Finally, we compute the disagreement between $\egwNR$ and $\egwBF$
  \begin{equation}
    \label{eq:delta_egw}
    \Delta \egw = \egwNR - \egwBF
  \end{equation}
\end{enumerate}
$\Delta\egw$ acts as a measure of the disagreement of the eccentricities between
the two methods of assigning eccentricity to a given NR simulation. One can
also think of it as a measure of disagreement of the eccentricities of an NR
waveform and the best-fit model waveform for a given waveform model.

\subsubsection{Loss Functions}
\label{sec:loss_functions}
One of the crucial part of finding $(\fBF, \eBF)$ is to use an appropriate
LF. In this work, we consider the following LFs:

\begin{enumerate}
\item {\itshape Mismatch:} The mismatch $\mm$, defined in equation \eqref{eq:mismatch},
  between the NR and the \teob waveform for a given initial frequency and eccentricity, is one of most used LF to measure the disagreement between two given GW waveforms. The mismatch also depends on the noise sensitivity $S_n(f)$ of the detector. Since the detectors are sensitive only within a particular window of frequencies,
  the mismatch only carries information from the relevant part of the signals given a particular detector, and is thus geared to practical applications rather than theoretical comparison of models. A "White Noise" mismatch, \ie a flat $S_n(f)$, on the other hand, is suitable for theoretical comparisons of waveform modes below the Nyquist frequency for a given sampling rate.  We considered two representative cases, namely the "White Noise" mismatch $\mathcal{M}_{W}$, \ie a flat $S_n(f)$, and the aLIGO-like mismatch $\mathcal{M}_{L}$. We only emphasize analyses dealing with the former and show some insight about the $LF\sim \mathcal{M}_L$ in Appendix \ref{subsec:results_details}, since the two different mismatch weightings led us to similar conclusions.

\item {\itshape $L_2$-norm of $\h_{22}$ residuals}  The match intervening in equation \eqref{eq:mismatch} is a correlation coefficient in Fourier space but not a distance, \ie derived from a norm itself defined from the inner product of the vector space.
  One can therefore use a more direct measure of the differences between the
  two waveforms by computing in time domain\footnote{for time domain waveforms, as it it the case in the scope of this paper} the normalized $L_2$ error
  
  \begin{equation}\label{eq:L2_h}
  \begin{aligned}
    \hL(f_0, e_0) &= \frac{\sqrt{\int_{t_1}^{t_2} \left|\h_{22}^{\text{NR}} -
        \h_{22}^{\text{TEOB}}(f_0, e_0)\right|^2\dd t}}{\sqrt{\int_{t_1}^{t_2}
      \left|\h_{22}^{\text{NR}} \right|^2\dd t}}\\
     &\equiv \sqrt{\|\Delta \h_{22}\|/\|\h_{22}^{\text{NR}}\|^2} \equiv \| \delta \h_{22}\|.
  \end{aligned}
\end{equation}
  We will note, for simplicity, $ \hL(f_0, e_0) = \|\delta \h_{22}\|$, also emphasizing that this scalar is a distance in time-domain functions and not a correlation measure of Fourier space vectors as the mismatch.
  This LF is at the core of works such as \cite{Blackman:2017dfb, Islam:2021mha,Islam:2022laz, Islam:2024tcs, Rink:2024swg}, notably to fit surrogate parametric models to reference NR simulations. 
\item {\itshape $\Delta \egw$:} In the cases where the waveforms are long
  enough, we can use $\Delta \egw$, defined in equation \eqref{eq:delta_egw}, in the LF to obtain ($\fBF, \eBF$). This is particularly important in finding the best
  matched model waveform for constructing a hybrid eccentric waveform, where
  the full inspiral-merger-ringdown waveform is built by smoothly stitching a model waveform (for inspiral part) and an NR waveform (for post-inspiral part).
  However, one should be careful about choosing the best matched waveform based
  solely on the the basis of minimal $\Delta \egw$ without ensuring that this waveform also gives a reasonably small $\mm$. 
  Ideally the $\mm$ associated
  with the minimized $\Delta\egw$ should be $< 1\%$. Therefore in
  Sec.~\ref{sec:degw_as_lf}, we will also address the question of whether
  minimizing $\Delta\egw$ also ensures  $\mm < 1\%$, without additional calibration.

\item {\itshape $C_\lambda$:} Finally, following the discussion in above,
  we study whether a combination of $\Delta\egw$ and $\mm$ would serve as a
  better LF than using only one of them to ensure both a small $\Delta\egw$ and
  a small mismatch, $\mm < 1\%$.
  \begin{equation}
    \label{eq:c_lambda}
    C_\lambda = \lambda |\Delta \egw| + (1-\lambda) \mm,
  \end{equation}
  where $\lambda$ is a scalar value in the range $[0, 1]$ and the $\Delta \egw$ term can be measured on a single frequency bin or  computed from the average difference over multiple bins in the inspiral phase.
\end{enumerate}

\subsubsection{LF minimization strategy}
\label{sec:LF_minimization_strategy}

\begin{figure}
    \includegraphics[width=0.49\textwidth]{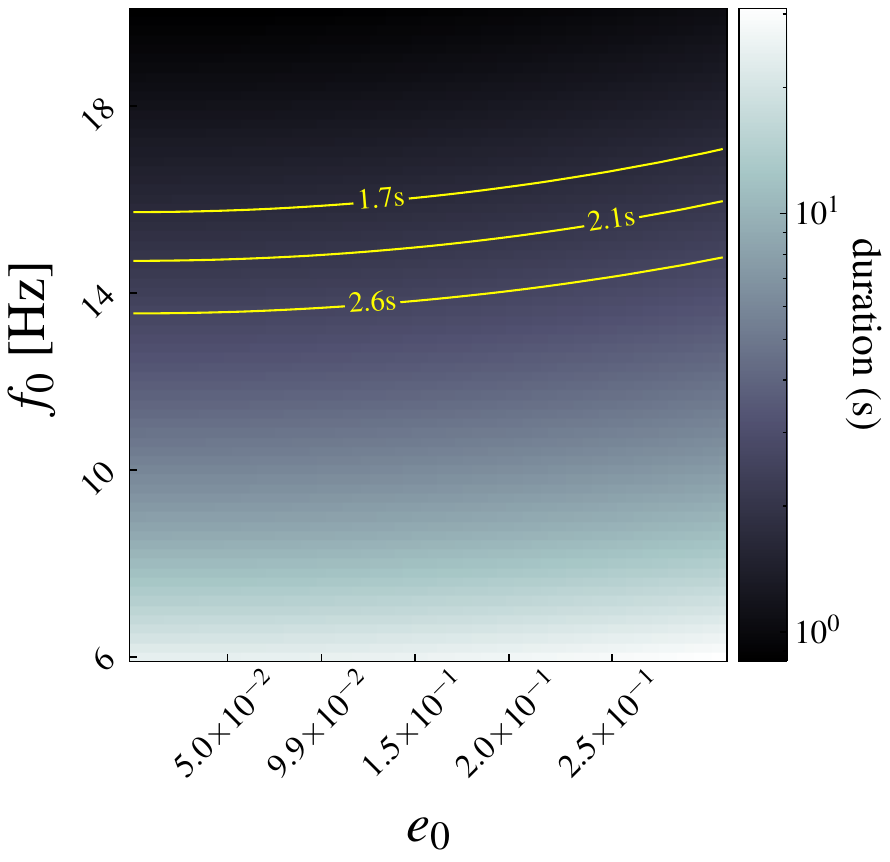}
    \caption{Duration of \teob varying on $(e_0,f_0)$ with the fiducial physical units chosen in this work. The yellow bounds above indicate the durations, with the same physical units, of the shortest and longest SXS simulations at our disposal. Physical quantities correspond to total binary mass $M = m_1 + m_2 = 60 M_{\odot}$ and luminosity distance $D_L = 400$Mpc.}
    \label{fig:duration_grid}
\end{figure}

An eccentric system requires two eccentricity parameters---eccentricity and
mean anomaly at a given reference frequency---to accurately describe
it. However, in our version of \teob, the mean anomaly parameter is fixed to a
certain value, and it is not a free parameter. Therefore, to effectively
explore the same parameter space, we vary the reference frequency. These two approaches are equivalent since we have only two additional independent parameters to vary for eccentric systems \cite[\eg][]{Knee:2022hth}. 
For all SXS waveforms, we estimate the duration of the inspiral as the time between the global maxima of $|A_{22}(t)|$ and the time at the start of the waveforms, for the physical units determined by our fiducial component masses and luminosity distance.
We show in Figure \ref{fig:duration_grid} the waveform duration of \teob in $(e_0, f_0)$ for the same physical units. We will use these durations to initialize the range of \teob parameter searches.

\begin{figure*}
    \includegraphics[width=0.95\linewidth]{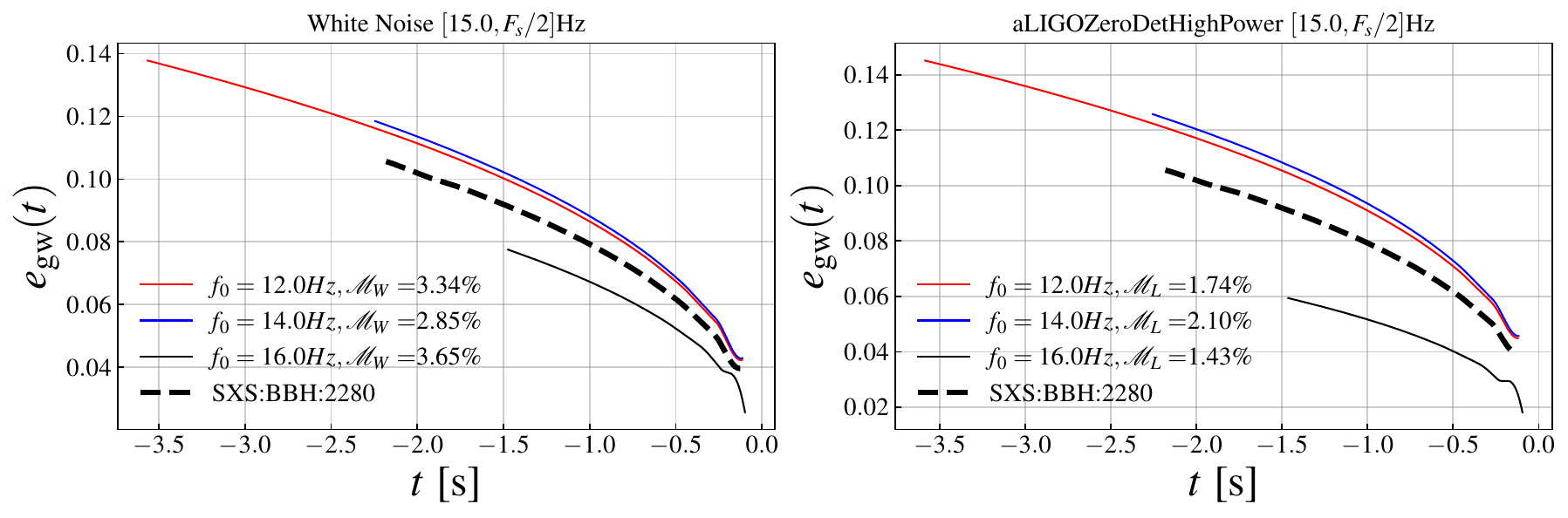}
    \caption{Minimization of the mismatch from equation \eqref{eq:mismatch}, for three cases of fixed reference frequency $f_0$, between \teob waveforms and some reference eccentric NR template, when $S_n$ is White Noise (left panel) or representative of aLIGO (right panel). Physical quantities correspond to total mass $M = m_1 + m_2 = 60 M_{\odot}$ and luminosity distance $D_L = 400$Mpc. This example illustrates the need to perform 2D searches over $(f_0, e_0)$, rather than fixing $f_0$ and searching over $e_0$ in our case.}
    \label{fig:1D_badidea}
\end{figure*}

As preliminary, we illustrate the need to vary both $e_0$ and $f_0$. While \cite{Joshi:2022ocr} fixed $f_0$ to fit waveform templates to NR sets by varying $e_0$ (in the context of studying the impact of higher $\ell$ modes), we show in Figure \ref{fig:1D_badidea} that in our study case fixing $f_0$ to some values would not result in neither satisfying $\egw$ agreements nor sub-percent mismatches in general. Indeed, we see in this plot that 
\begin{itemize}
    \item we must also vary $f_0$ in addition to $e_0$ to minimize the mismatch (or any other LF a-priori) with some reference eccentric NR template.
    \item In addition to $\egw$ from our SXS set that gives us an approximate idea of the parameter search range, we must also use NR data to find an appropriate search range for $f_0$.
    \item The noise PSD $S_n$ will have an impact on the best-match waveform. Even though this question was part of our initial motivations, we only show results using two different mismatches in Table \ref{table:summary}, since our overall prescriptions are the same.
\end{itemize}

\begin{figure}
    \includegraphics[width=0.49\textwidth]{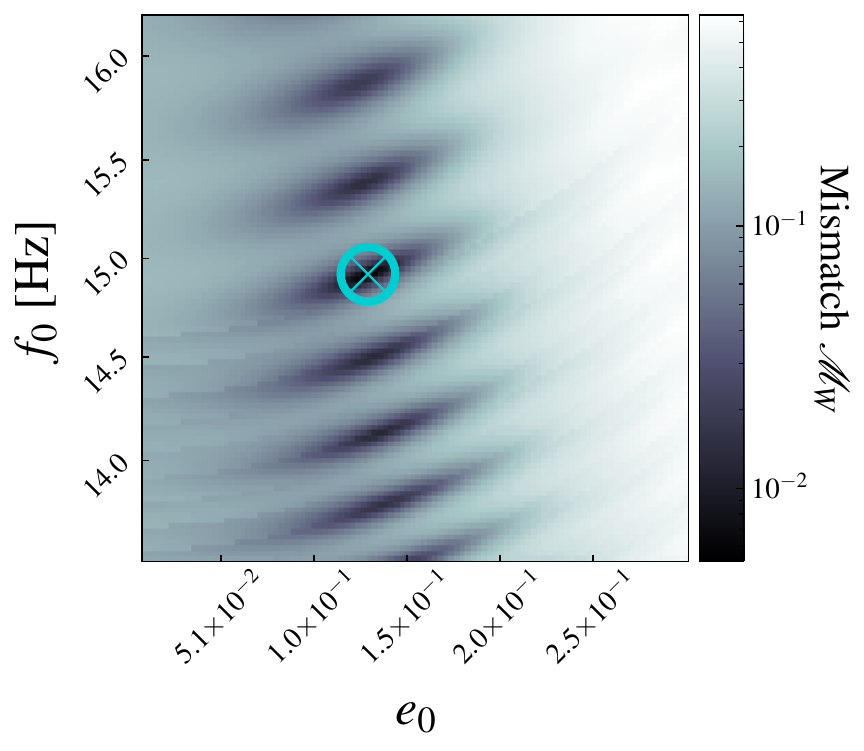}
    \includegraphics[width=0.49\textwidth]{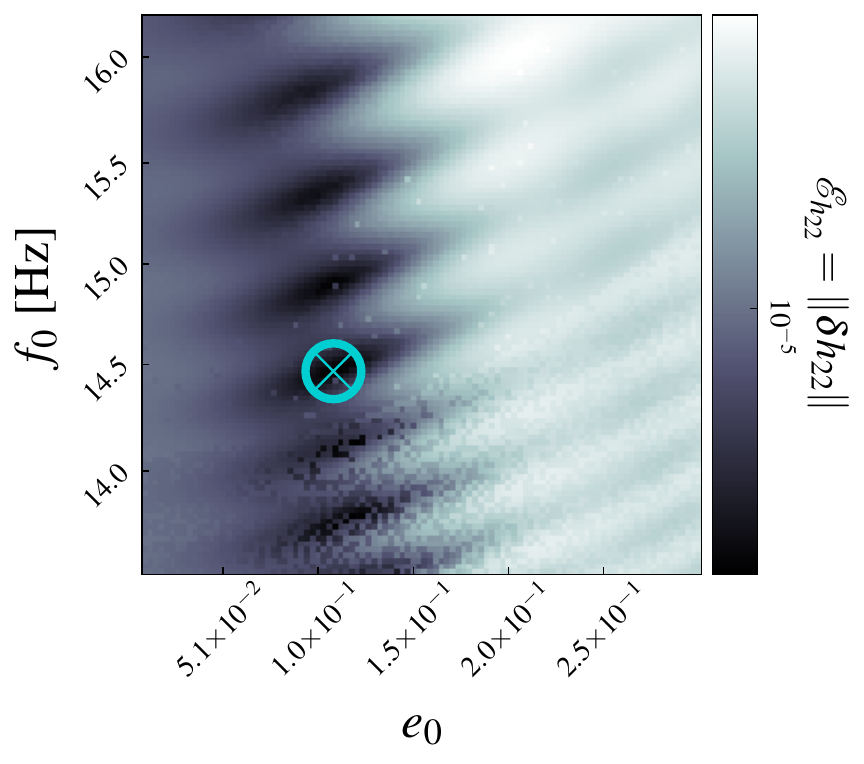}
    \caption{LFs on coarse grid from \texttt{SXS:BBH:2285}. We indicate the coordinates of LF minima on the grid with cyan circles. \underline{Top panel:} Mismatch with flat $S_n(f)$ referred to as $\mm_W$.
    \underline{Bottom panel:} $L_2$-norm of the \mode mode residuals referred to as $\|\delta \h_{22}\|$ (equation \ref{eq:L2_h}. Physical quantities correspond to total binary mass $M = m_1 + m_2 = 60 M_{\odot}$ and luminosity distance $D_L = 400$Mpc.}
    \label{fig:LFgrid_mismatch}
\end{figure}

The most crucial step in obtaining the ($\fBF, \eBF$) pair is finding an robust strategy to minimize the LF over ($f_0,
e_0$). This is because the LFs based on waveform modes have multiple local valleys making it hard to find the global minimum. We illustrate the shapes of $\|\delta h_{22}\|$ and $\mm_W$ in the $(e_0, f_0)$ plane, determined for each SXS simulation metadata, in Figure \ref{fig:LFgrid_mismatch}.

 In order to initialize the parameter search scope from NR data, one can use, as in \cite{Islam:2021mha}, estimates of the mean anomaly and eccentricity from the NR waveforms. Since our \teob implementation allows to vary only eccentricity and reference frequency, we will use measures of $\egw$ and the waveform durations instead.
\begin{enumerate}[label=(\arabic*)]
\item \label{it:ini_from_NR_meta} 
  For all SXS waveforms, we find appropriate values of the initial \teob parameters by requiring the duration of the inspiral to be similar. Moreover, we have already measured $\egw (f)$ in the left panel of Figure \ref{fig:sxs_egw}, which gives us appreciations of eccentricity range in the early part of the inspiral.
\item \label{it:find_ini_for_same_length}
  Next, we perform a coarse grid
  search over $(f_0, e_0)$ in the range determined by the previous step \ref{it:ini_from_NR_meta}, as shown in Figure \ref{fig:LFgrid_mismatch} for instance.
\item \label{it_nelder_mead} Find a small rectangular box around the rough parameters obtained in step \ref{it:find_ini_for_same_length} and use a Nelder-Mead search given some LF to get the refined pair $(\fBF, \eBF)$.
\end{enumerate}

The coarse grid computation to initialize a Nelder-Mead search is in the spirit of \cite{Knee:2022hth}, although this work used a much thinner spacing on the waveform parameters coarse grid search.
For this reason, at the cost of more brute force computations, we double-checked our $(\eBF, \fBF)$ pairs from step \ref{it_nelder_mead} by initializing $\mathcal{O}(100)$ Nelder-Mead searches at each $\tilde{e_0^i}$ that minimizes the LF for each frequency bin $f_0^i$ of the coarse grid (\eg the vertical axes of both panels of Figure \ref{fig:LFgrid_mismatch}). While these redundant searches did not modify the conclusions, we obtained slightly lower LF values for some cases, thus allowing for more confidence in our minima.

In the next section, we comment our LF minimizations on our $10$ SXS simulations and study the correspondence with the measured $\egw$.
\section{LF minimization and eccentricity}
\label{sec:Results}
To answer the aspects of this study, we first show the resulting $\egw$ evolutions from the best-fit \teob waveforms in Section \ref{subsec:LF}. Then, we discuss the results and propose prescriptions to use $\egw$ when fitting eccentric waveforms in Section \ref{subsec:interpretation}. All the plots stem from the best-fit results detailed in Table \ref{table:summary}, and the reader can jump to Figure \ref{fig:money_plot} as a companion to our conclusions and prescriptions.

\subsection{Fitting waveforms with Loss Functions}\label{subsec:LF}
\begin{figure*}[htb]
  \centering 
  
  \includegraphics[width=0.98\columnwidth]{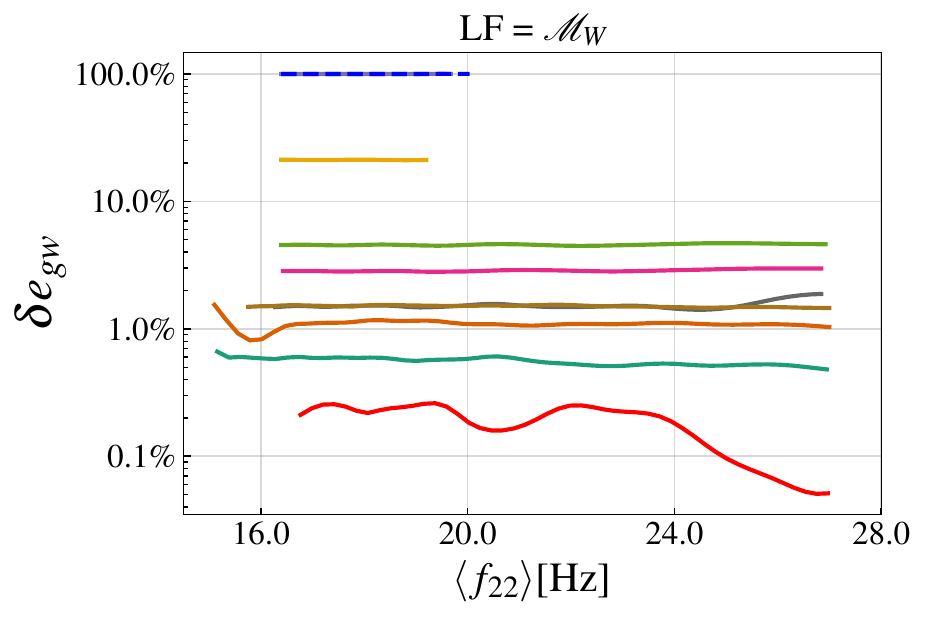}
  \includegraphics[width=0.98\columnwidth]{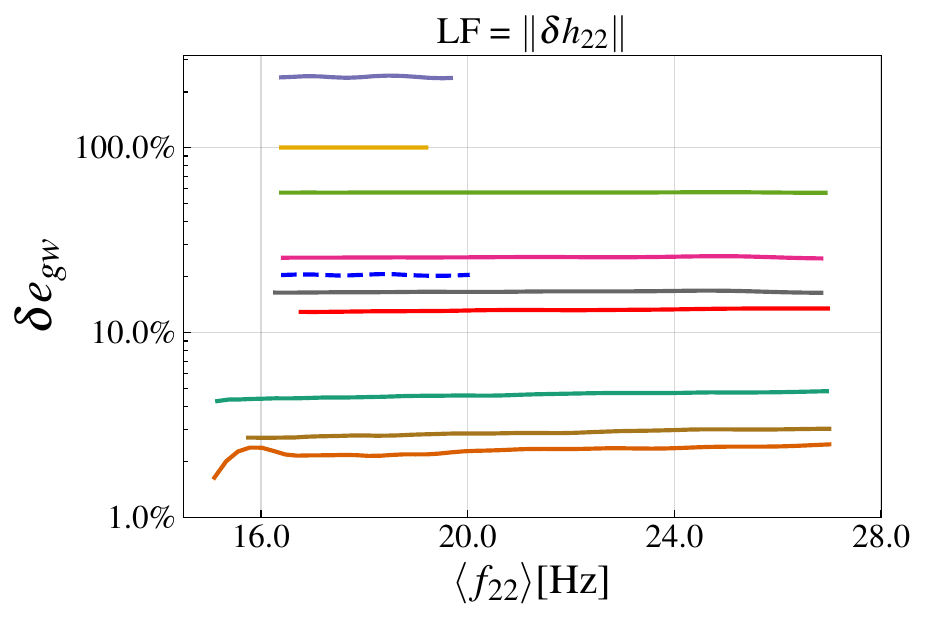}

  \includegraphics[width=0.98\columnwidth]{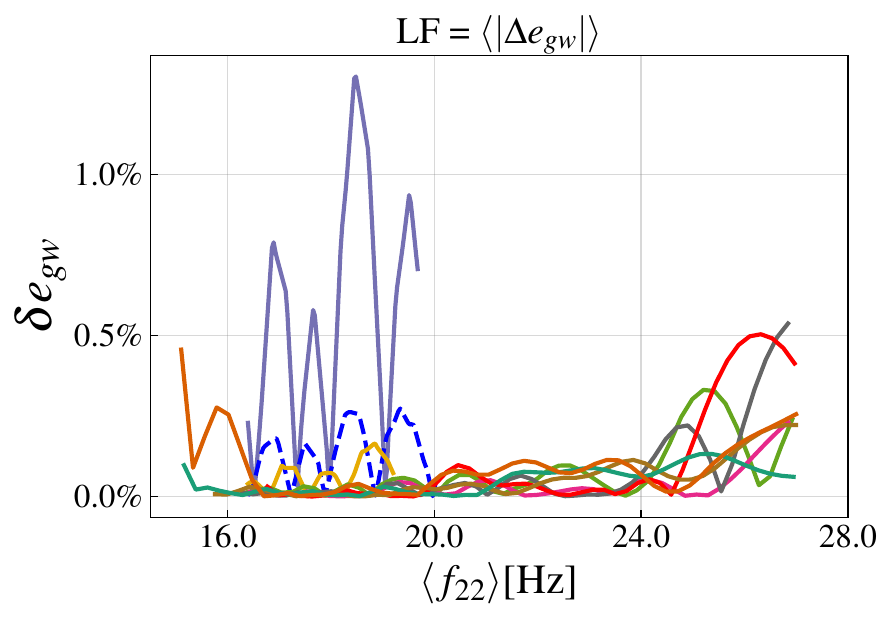}
  \includegraphics[width=0.98\columnwidth]{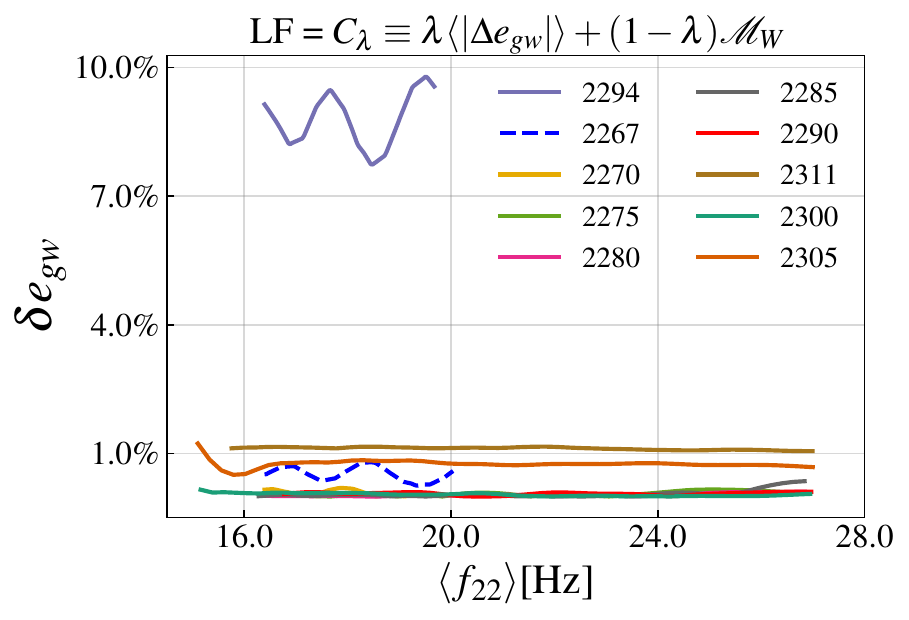}
\caption{Fractional $\egw$ differences from \teob generated from $(\eBF,\fBF)$ with respect to each SXS simulation from Table \ref{table:SXSdata}. Each of the four panel corresponds to one LF of Section \ref{sec:loss_functions}. In the top left panel, the curves for \texttt{SXS:BBH:2267} and \texttt{SXS:BBH:2294} illustrate that the $\mm_W$ minimization favors the quasi-circular regime of \teob, resulting in $\delta \egw \sim 100 \%$ (up to numerical errors). The bottom right panel corresponds to $\lambda =0.3$. Physical quantities correspond to total mass $M = m_1 + m_2 = 60 M_{\odot}$ and luminosity distance $D_L = 400$Mpc. Note that we only show a fraction of the measurable $\egw$ frequency range, until $f=28$Hz, since the error $\delta \egw (f)$ becomes less relevant as we get closer to merger time.}
\label{fig:LFs_degw}
\end{figure*}

As a starting point, Figure \ref{fig:LFs_degw} presents the fractional $\egw$ differences

\begin{equation}\label{eq:frac_degw}
    \delta \egw (f)\equiv \left|\frac{\egw^{\text{TEOB}}(f)}{\egw^{\text{NR}}(f)} -1\right|,
\end{equation}
in frequency domain, during the inspiral phase, measured after finding the best-fit parameters $(\eBF,\fBF)$ for each LF introduced in Section \ref{sec:loss_functions}. We stress that the $\egw$-based LF is calculated with the absolute difference $\Delta \egw$ - more appropriate as a LF especially for small eccentricity cases -  while we will show the fractional difference of equation \eqref{eq:frac_degw} in some plots. To ease the discussions, we will designate \texttt{SXS:BBH:2294}, \texttt{SXS:BBH:2267} and \texttt{SXS:BBH:2270} from Table \ref{table:SXSdata} as the lowest eccentricity simulations from now on.

\subsubsection{LF = $\mm$}\label{sec:mismatch_as_lf}
In the top left panel of Figure \ref{fig:LFs_degw}, we observe that the highest $\egw$ fractional differences correspond to to the low eccentricity cases. In particular, \texttt{SXS:BBH:2294} and \texttt{SXS:BBH:2267} are best-fit with $e_0=0$ using the mismatch. Except for \texttt{SXS:BBH:2270}, all best-match waveforms fall below $\degw \leq 5\%$ in \textit{fractional} disagreement. 
Nonetheless, we will have to address two points in the next section. 
\begin{enumerate}
    \item the quasi-circular waveform can be enforced by the mismatch minimization for small eccentricities (in this case \teob)
    \item The mismatch and $\egw$ disagreement are not one-to-one correspondent, \ie the lowest mismatches do not result in better $\egw$ agreements.
\end{enumerate}
We note that using a different $S_n(f)$, in the aLIGO case, also led to a quasi-circular best mismatch, although not for the same cases (\texttt{SXS:BBH:2294} in Table \ref{table:summary}).

\subsubsection{LF = $\hL$}
\label{sec:hL_as_lf}
The top-right panel of Figure \ref{fig:LFs_degw} presents the $\egw$ measurements with respect to SXS simulations after minimizing for equation \eqref{eq:L2_h}, \ie the $L_2$-norm in time domain of aligned \mode modes. Overall, our minimization strategy from Section \ref{sec:LF_minimization_strategy} results in poorer $\egw$ agreements than with the mismatch, especially for small eccentricities. We also note that this LF also resulted in a quasi-circular best-fit waveform in one case (\texttt{SXS:BBH:2270}) The poor agreements in comparison with the mismatch LF could be counter-intuitive since the \mode mode residuals is a proper distance, whereas the mismatch is a correlation coefficient, \ie a measure of the cosine of the angle between the vectors of Fourier coefficients  "normalized" on the hypersphere. Moreover, as we will illustrate in more details in Section \ref{subsec:interpretation}, $(\eBF, \fBF)$  from $\|\delta \h_{22}\|$ minimization lead to relatively high mismatch best-fit waveforms, typically with $\mm_W$ of a few percents (Table \ref{table:summary}). 

\subsubsection{LF = $\Delta \egw$}
\label{sec:degw_as_lf}
\begin{figure}
    \includegraphics[width=0.49\textwidth]{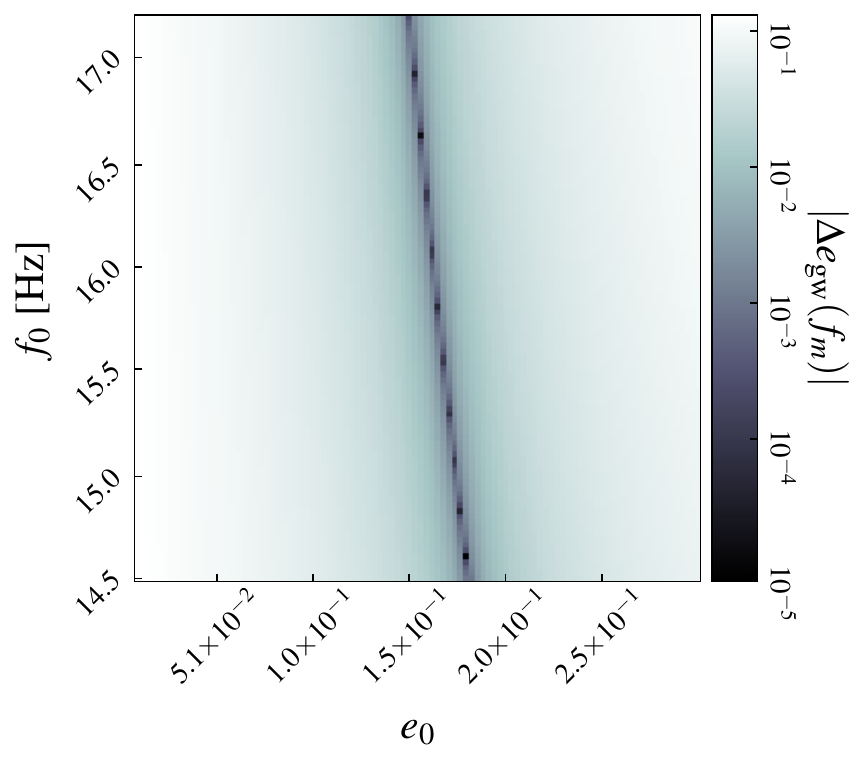}
    \caption{Illustration on \texttt{SXS:BBH:2290} of a LF derived from $\Delta \egw$, here $|\Delta \egw(f_m)|, f_m=18$Hz. Physical quantities correspond to total binary mass $M = m_1 + m_2 = 60 M_{\odot}$ and luminosity distance $D_L = 400$Mpc.}
    \label{fig:degw_grid}
\end{figure}

In Figure \ref{fig:degw_grid}, which shows $|\Delta \egw|$ on the coarse $(e_0,f_0)$ grid, we observe that the valley extending in the $f_0$ direction is significantly narrower than in the case of the mismatch, although oscillations are still present. This valley also crosses local minima and maxima of the LFs of Figure \ref{fig:LFgrid_mismatch}.
In the bottom-left panel of Figure \ref{fig:LFs_degw}, we show that as expected, the $\egw$ agreement is the lowest when fitting directly on $\Delta \egw$. What does not appear on this plot, however, is that the best-fit waveforms have high mismatches, up to $\sim 10\%$ (Table \ref{table:summary}) for the highest eccentricity cases of Table \ref{table:SXSdata}.
Thus, as one could expect from the fact the $\egw$ only invokes the phase evolution $\Phi_{22}$ in equation \eqref{eq:22mode}, this LF is not appropriate on its own.

\subsubsection{LF = $C_{\lambda}$}\label{sec:c_lamba_as_lf}
Our motivation for $C_{\lambda}$ is twofold. First, for long NR waveforms that have a sufficient number of orbits, a simple linear combination of two LFs, one of them computed with $\egw$, is an easy solution to incorporate the standardized eccentricity definition. In addition, this hybrid LF allows by construction to deal with cases where mismatch minimization would enforce quasi-circular best-fit waveforms by controlling the level of $\egw$ disagreement. While such cases primarily serve as a diagnosis of the limitations of the waveform in use, it is also important to characterize eccentricity in reference simulations with quantified $\mm$ and $\Delta \egw$ requirements which would, in turn, guide choices for $\lambda$. The case for finding the hyperparameter $\lambda$ is beyond the scope of our this paper, thus we will simply present results from particular values. Unless stated otherwise, we present results for $\lambda = 0.3$, as in Figure, which allowed not to assign quasi-circular waveforms by minimization in the smallest eccentricity cases and to obtain best-fit waveforms with sub-percent mismatches.
The bottom-right panel of Figure \ref{fig:LFs_degw} demonstrates that the lowest eccentricity SXS cases are now fitted with an eccentric waveform with both sub-percent-mismatches (see Table \ref{table:summary} and Figure \ref{fig:money_plot}) and - except for the smallest eccentricity \texttt{SXS:BBH:2294} - percent level best-fit $\egw$ agreements.

\subsection{Interpretation and prescriptions}\label{subsec:interpretation}

\begin{figure*}
\includegraphics[width=0.49\textwidth]{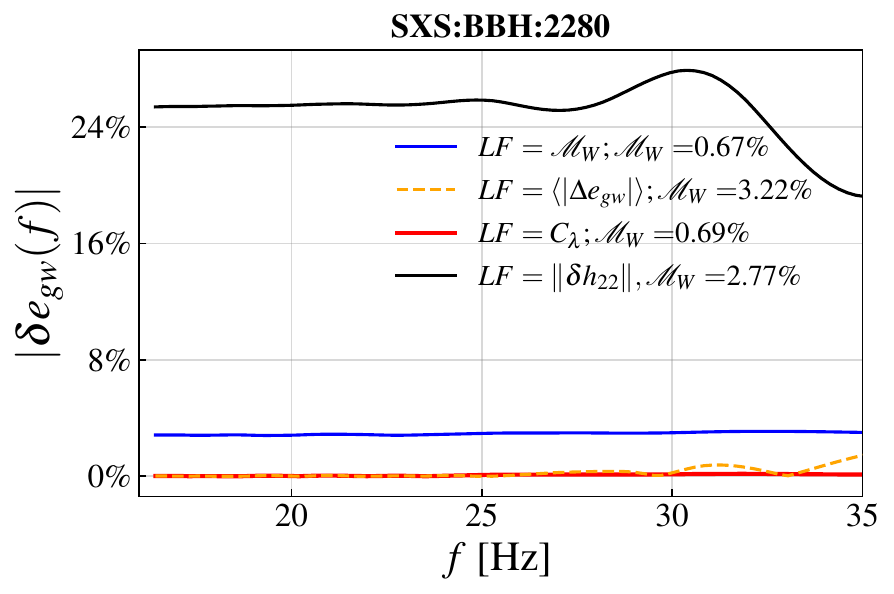}
\includegraphics[width=0.49\textwidth]{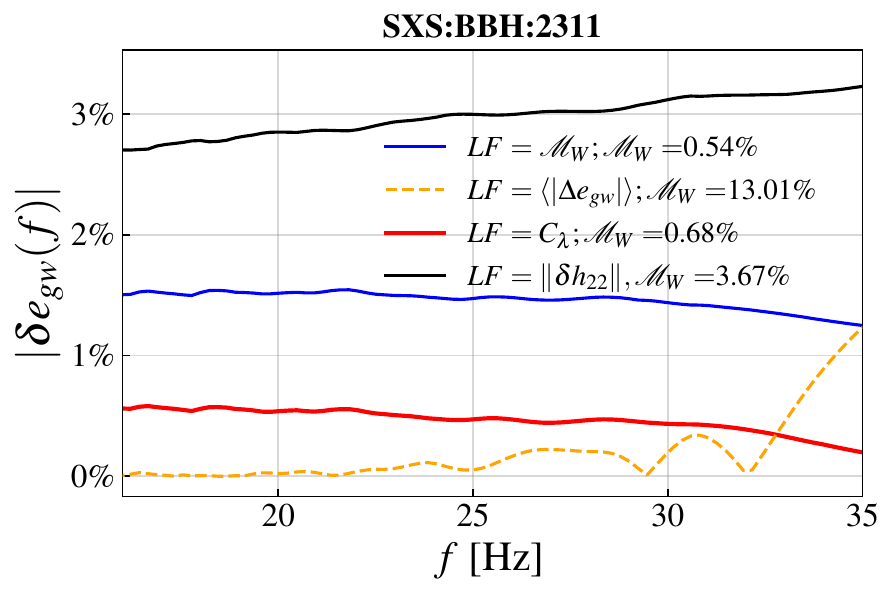}
  \caption{Frequency evolution of the fractional difference $\delta egw$ of the best-fit \teob waveforms for our LFs from Section \ref{sec:loss_functions}, with respect to the measurements of the fixed SXS waveforms (one example in each panel.) The loss function $C_{\lambda}$ has $\lambda = 0.3$ on the left panel and $\lambda = 0.5$ on the right panel, where for instance $\lambda=0.3$ results in $|\delta \egw|$ slightly above $1\%$ in this frequency range. Physical quantities correspond to total mass $M = m_1 + m_2 = 60 M_{\odot}$ and luminosity distance $D_L = 400$Mpc.}
  \label{fig:degw_insight}
\end{figure*}

 Figure \ref{fig:LFs_degw}, only tells us about the $\egw$ agreement without actually informing about the similarity of the best-fit waveforms with their NR counterpart. Therefore, we show the mismatches of the best-fit waveforms for different LFs in addition to the $\egw$ agreement evolution for two representative SXS examples in Figure \ref{fig:degw_insight}, namely \texttt{SXS:BBH:2280} and \texttt{SXS:BBH:2311}. As previously stated, fitting on $\Delta \egw$ (orange lines) can generate best-fit waveforms with high mismatch values. Moreover, we indicate our $\h_{22}$ minimization results and emphasize, in addition to the poorer $\egw$ agreements, the higher mismatches $\mm_W \sim 3\%$ in theses examples for this LF. On the left panel, for \texttt{SXS:BBH:2280}, the convex loss $C_{\lambda}$, at the cost of an additional mismatch of $\approx 2 \times {10^{-4}}$, improves the $\egw$ agreement from $\sim 3\%$ to $\sim 0.01\%$. Of course, the practical relevance of $C_{\lambda}$ would depend on thresholds for the mismatch given a waveform. On the right panel, minimizing $\mm_W$ already gives a sub-percent $\egw$ agreement in the measured band.

\begin{figure}
    \includegraphics[width=0.5\textwidth]{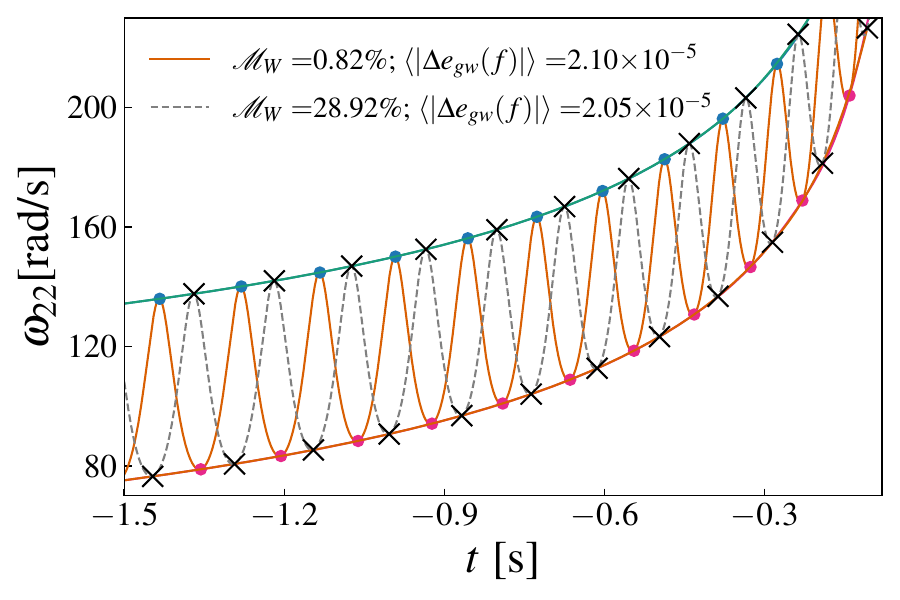}
    \caption{Illustration on \texttt{SXS:BBH:2300} of what we refer to as $\egw$ "overfitting" when $LF \sim \Delta \egw$. From an arbitrarily close $\sim \mathcal{O}(10^{-5})$ $\egw$ fractional error, we explored the $(e_0, f_0)$ parameter space to find the worst and best mismatch possible, which correspond to $\omega_{22}$ interpolants having same opposite and same phase as the NR phase evolution, respectively. Physical quantities correspond to total mass $M = m_1 + m_2 = 60 M_{\odot}$ and luminosity distance $D_L = 400$Mpc.}
    \label{fig:overfitting_visu}
\end{figure}

Regarding our conclusion on LFs derived from $\egw$, Figure \ref{fig:overfitting_visu} illustrates what we refer to as the "degeneracy" with the mismatch, which measures waveform similarity. As we previously emphasized, $\egw$ considers only the phase evolution of the waveform. Additionally, its computation requires interpolation through the times of apastron and periastron. Therefore, an infinity of $(e_0,f_0)$ pairs can result in $|\Delta \egw|$ below a required threshold. A difference of the order of $10^{-5}$ in $|\Delta \egw|$ can either give a sub-percent mismatch or a mismatch of $\sim 30\%$ as in the example of Figure \ref{fig:overfitting_visu} where the carefully chosen extremes have opposite phases but the same interpolants through the local extrema of $\omega_{22}$.

\begin{figure*}
    \centering \includegraphics[width=1.1\linewidth]{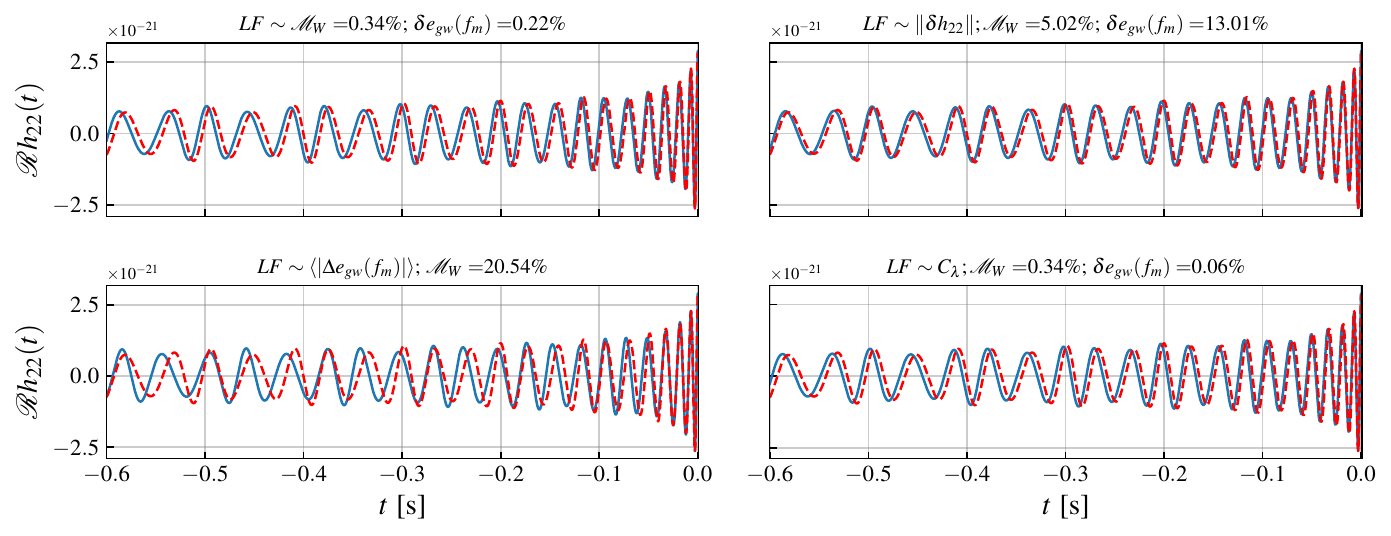}
    \caption{Time evolution of the real part of the \mode mode of the best-fit \teob waveforms for \texttt{SXS:BBH:2290}. The strains are coaligned, \ie all the waveforms are shifted so that $t_m=0$ and $\Phi_m=0$ at merger time. Coalignment \underline{does not} correspond to the time and phase shifts maximizing the match from equation \eqref{eq:mismatch}.  We indicate at the top of each panel, for each LF from Section \ref{sec:loss_functions}, the mismatch  and the $\egw$ difference of the best-fit waveform with the NR counterpart.}
    \label{fig:h22vizu_LFs}
\end{figure*}

\begin{figure}
    \includegraphics[width=0.49\textwidth]{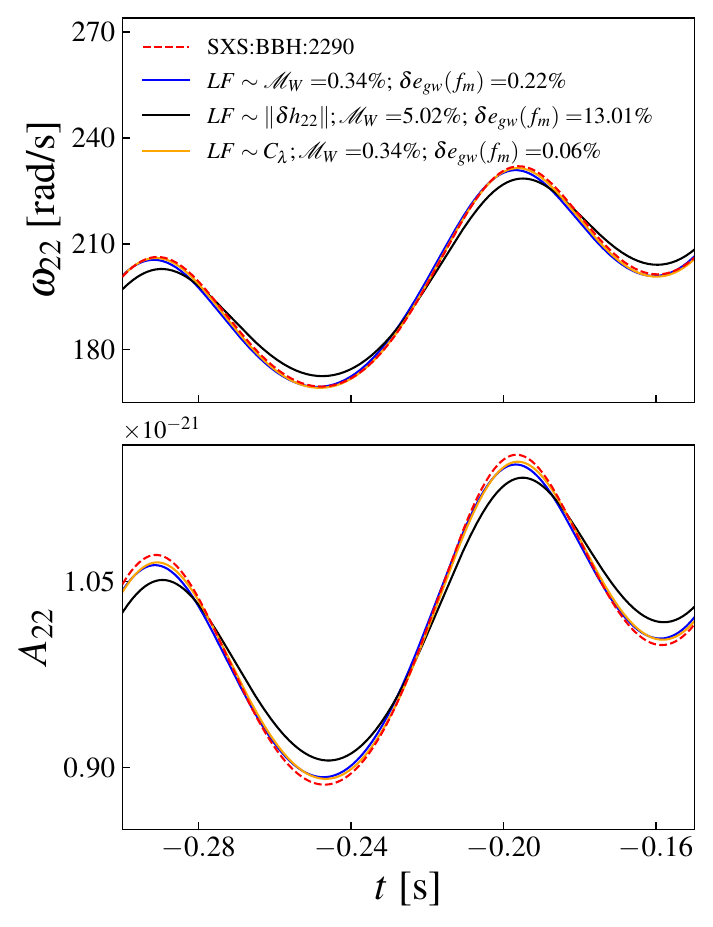}
    \caption{Phase and amplitude evolution of the \mode mode of the best-fit \teob waveforms in the case of \texttt{SXS:BBH:2290}, where the waveforms have phase $\Phi_m=0$ at merger time $t_m = 0$. This plot is a companion to Figure \ref{fig:h22vizu_LFs}. Physical quantities correspond to total binary mass $M = m_1 + m_2 = 60 M_{\odot}$ and luminosity distance $D_L = 400$Mpc.}\label{fig:22mode_example}
\end{figure}

To explain our findings, particularly about the poor performances of the $L_2$-norm from equation \eqref{eq:L2_h}, we present the time evolution of the \mode mode for a specific SXS example, comparing it with the the best-fit \teob waveforms in Figures \ref{fig:h22vizu_LFs} and \ref{fig:22mode_example}. We chose to show in Figure \ref{fig:h22vizu_LFs} the coaligned waveforms rather than those shifted in time and phase as to maximize the match. Since all the strains have the merger time at $t=0$ and a null phase, this allows us to visually examine the actual computation for LF = $\hL$ in time domain. While the red (SXS) and blue (\teob) strains look the most similar in the upper-right panel, the mismatch is important ($\mm_w  = 5\%$) and so is the $\egw$ difference, as we can see in the earlier part of the inspiral on the plot. Figure \ref{fig:22mode_example} actually emphasizes the phase and amplitude evolution differences for the same \mode mode as in the previous Figure \ref{fig:h22vizu_LFs}. We argue that minimizing the difference of sampled $\h_{22}$ vectors in time domain does not allow precise control over which part of the waveform is prioritized until merger nor does it ensure which frequency band should be trusted for a specific eccentric waveform model. There is no guarantee that minimizing equation \eqref{eq:L2_h} over a given time interval, \textit{on average}, will ensure sufficient correlation of Fourier modes, as our examples suggest. This problem can be mitigated using the mismatch, though at the cost of relying on a specific $S_n(f)$.

\begin{figure*}
    \includegraphics[width=0.75\linewidth]{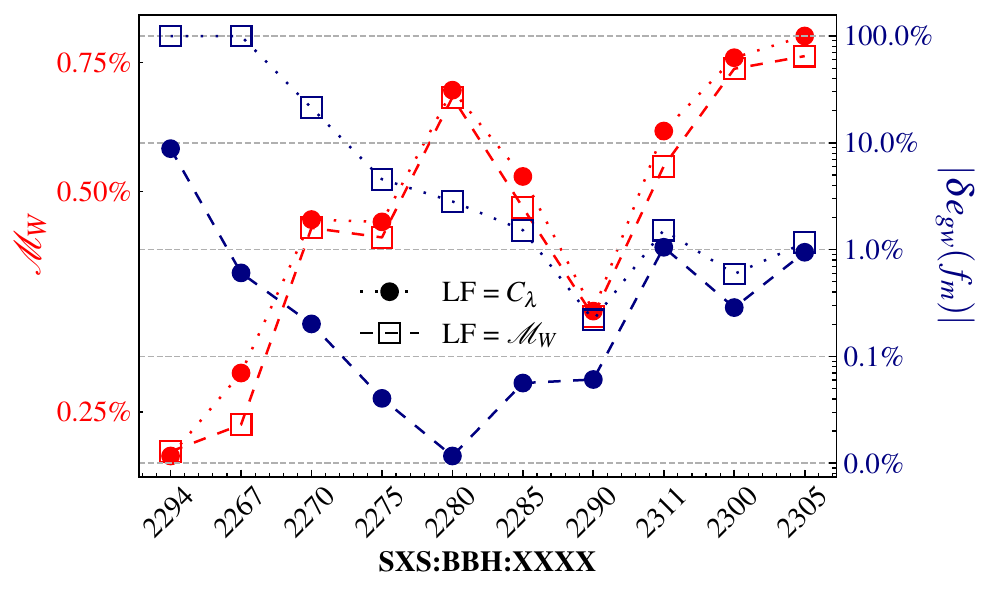}
    \caption{Cross examination of the mismatch LF $\mm_W$ with flat $S_n$ (equation \eqref{eq:mismatch}, squares) and $C_{\lambda}$ LF with $\lambda = 0.3$ (\eqref{eq:c_lambda}, circles). minimization results. Whichever LF is minimized, we show the mismatch of the best-fit \teob waveform in red (left $y$-axis) and fractional $\egw$ difference in indigo blue (right $y$-axis). The 10 SXS identifications on the $x$-axis are sorted by increasing $\egw (f_m=18Hz)$ from Table \ref{table:SXSdata}. We emphasize that by using $LF \sim C_{\lambda}$, at the price of a higher mismatch than with $LF \sim \mm_W$, depending on $\lambda$, we enforce eccentricity agreement by construction. For the smallest eccentric simulations \texttt{SXS:BBH:2294} and \texttt{SXS:BBH:2267}, $C_{\lambda}$ allows to assign eccentricity to the NR strain rather than preferring the quasi-circular approximant with the mismatch.}
    \label{fig:money_plot}
\end{figure*}

Finally, In Figure \ref{fig:money_plot}, we sort the SXS simulations from Table \ref{table:SXSdata} along the $x$-axis in order of increasing $\egw$, and, when fitting \teob to these simulations using either the mismatch or the mismatch combined $\Delta \egw$ ($LF \sim \mm_W$ or $LF \sim C_{\lambda}$, respectively), we display the corresponding fractional eccentricity error $\delta \egw$ and mismatch from the \teob waveform using $(\eBF, \fBF)$. On this Figure, we observe that all the $\hBF$ strains resulting from either of these two LFs have a mismatch $\mm_W \leq 1\%$. Our fiducial case is $\lambda = 0.3$; the case $\lambda = 0.5$ led to mismatched higher than $1\%$ for the lowest eccentric cases and $\lambda = 0.1$ resulted in a quasi-circular \teob best-fit for \texttt{SXS:BBH:2267}. We argue that, in practical cases (\eg fitting a surrogate model to a reference waveform or characterizing eccentric NR simulations with a model), the choice of $\lambda$ will be guided by specific match and eccentricity agreement requirements.
Then, we emphasize that when fitting our \teob implementation with only the mismatch for the small eccentricity SXS simulations (\texttt{SXS:BBH:2294} and \texttt{SXS:BBH:2267}), the minimization procedure described in Section \ref{sec:LF_minimization_strategy} unequivocally favors the quasi-circular approximation $e_0 = 0$.
The detailed values of Figure \ref{fig:money_plot} can be found in the corresponding colored entries in Table \ref{table:summary}.

Overall, we stress that we do not propose that $C_{\lambda}$ as a replacement for the mismatch, which ensures correlation of the waveform Fourier modes. Instead,  $C_{\lambda}$ provides a way to address "failure" cases, such as when the model at hand favors a quasi-circular approximant, as observed for \texttt{SXS:BBH:2294} and \texttt{SXS:BBH:2267}. Indeed, in the limit $\lambda \rightarrow 1$, we obtain an $\egw$-based LF that can be overfitted, allowing the mismatch to increase arbitrarily between the worst- and best-matched waveform for a given order of magnitude in $\Delta \egw$, as we illustrated in Figure \ref{fig:overfitting_visu}. In other words, we advocate to use a very simple LF like $C_{\lambda}$ to deal with model limitations while controlling the mismatch threshold with $\lambda$, as emphasized in Figure \ref{fig:money_plot}.
\section{Conclusion and remarks}
\label{sec:Conclusion_and_remarks}
The recent development of the gauge-independent measurement $\egw$ by the authors of \cite{Shaikh:2023ypz} motivated our study about LF minimization through numerical searches over the eccentricity parameters of a waveform model, with respect to a reference waveform. Our goal was to provide insight into the relationship between commonly used LFs, particularly the mismatch, and the resulting $\Delta \egw$, \ie the difference between the best-fit model and the fixed reference. Applications range from mapping eccentricity between waveform models, fitting fast surrogate models with the perspective or improving inference capabilities in the future, or characterizing eccentric NR simulations where the eccentricity is ill-defined in the GR regime from the initial conditions. 

We conducted numerical searches to minimize LFs, with the strategy from Section \ref{sec:LF_minimization_strategy}, on the eccentricity $e_0$ and reference frequency $f_0$ parameters of the EOB model \teob with respect to $10$ eccentric SXS simulations with $\egw$ ranging from $ \sim 10^{-3}$ to $\sim 0.2$ at $18$Hz. Our fiducial LFs are the mismatch from equation \eqref{eq:mismatch} - "White Noise", \ie with a flat $S_n(f)$ weighting -, the residuals in time domain of the \mode mode $h_{22}$ as per equation \eqref{eq:L2_h}, and LFs derived from the difference $\Delta \egw$. 
As expected - since $\egw$ utilizes only partial information from the dominant mode to determine interpolants through apastron and periastron times -  fitting to arbitrarily small $\egw$ differences can lead to high mismatches. Consequently, the best-fit waveform may not be physically meaningful as it does not represent the same binary system, even if the eccentricity evolution matches within the measured frequency range. Moreover, our strategy from Section \ref{sec:LF_minimization_strategy} with $h_{22}(t)$ residuals as the LF produces insufficiently matched waveforms ($\mm\sim$ a few percents) and poor $\egw$ agreements. We still report results for this LF in some plots and in the summary Table \ref{table:summary}.
When the LF is the mismatch, we can find $(\eBF, \fBF)$ pairs such that $\mm \sim 10^{-3}$ for most of the SXS set and closer to $\sim 10^{-2}$ for the highest eccentricity cases ($\egw \sim 0.2$). Moreover, the fractional difference $\delta \egw$ remains within a few percents for $\egw < 0.2$. Nonetheless, for our two less eccentric SXS simulations, minimizing $\mm$ can give quasi-circular $e_0=0$ best-fit waveforms. To circumvent this limitation - which may arise not only with \teob, but also more generally - we propose a linear convex combination of the mismatch and a (positive) LF derived from $\egw$. The parameter $\lambda$ controls the allowed deviation from the best $\mm$, which, in the ideal case, represents the LF we would recommend based on our findings for the aforementioned practical applications.

In addition, the recent study \cite{Bonino:2024xrv} would corroborate this prescription on the mismatch, although not in the context of numerical searches as in our work. The authors, after assigning eccentricities to their own NR simulations set through a mapping procedure of the initial conditions, find $\mm \sim 10^{-3}-10^{-2}$ (Figure 12 of current version) as a robustness metric rather than a LF. We argue that $C_{\lambda}$ is merely an alternate LF to deal with both parametric model limitations and mismatch requirements. A simple to implement LF like $C_{\lambda}$ will be particularly useful in the context of fitting surrogates to reference strains, among the practical applications that motivated our study. We leave new developments on fitting procedures invoking $\egw$ in addition to classic LFs, with numerical searches or other strategies as in \cite{Bonino:2024xrv}, to future studies.



\begin{acknowledgments}
We thank Vijay Varma and Leo C. Stein for useful discussions.
This work was supported by the National Research Foundation of Korea under grant No.~NRF-2021M3F7A1082056.
\end{acknowledgments}

\bibliography{References}

\appendix
\section{Additional Details }\label{app:insights}

\subsection{LF computations}\label{subsec:LF_details}
Below, we provide some additional details about our computations of the LFs introduced in Section \ref{sec:loss_functions} for the curious reader.

\begin{enumerate}
    \item We estimate the mismatch expressed in equation \eqref{eq:mismatch} with the \code{psd.filter.matchedfilter.match} method from the publicly available library \code{PyCBC} \cite{pycbc}. The lower integration bound is $\sim f_0 + 0.02$Hz for the waveform to start slightly before the band and the higher integration bound is the Nyquist frequency determined the the sampling rate in physical units. The advanced LIGO approximation in the $\mm_L$ cases - only shown in Table \ref{table:summary} - is the \code{aLIGOZeroDetHighPower} model from \code{pycbc.psd}
    \item To compute the $L_2$ loss in time domain from equation \eqref{eq:L2_h}, we coalign the waveforms and compute the numerical counterpart to equation \eqref{eq:L2_h} from the shared earliest time sample of the waveforms until merger time, which is defined by the maximum of the amplitude modulus $|A_{22}(t)|$.  To do so, we shift the waveform and NR data in time and phase so that they have both the same phase $\Phi_m = 0$ at merger time $t_m=0$. Then, to compute the LF using the same time vector for $\h_{22}^{\text{NR}}(t)$ and $\h_{22}^{\text{TEOB}}(t)$, we rely on linear interpolants.
    \item To calculate $\Delta \egw (f)$, we use linear interpolants in the common range $[f_{min}, f_{max}]$ within which we can measure $\egw$ for the waveform-NR pair. We compute either the difference measured at the specific bin $f_m=18$Hz, where we can measure $\egw$ consistently in all our cases, or the average difference $\langle| \Delta \egw| \rangle$ on frequency bins in the range $[f_{min}, 20Hz]$ where $f_{min}$ is the lowest frequency at which $\egw$ is definite after $f=\fBF$.
    \end{enumerate}
 
\subsection{LF minimization results}\label{subsec:results_details}
We present in Table \ref{table:summary} below a summary of the different LFs introduced in Section \ref{sec:loss_functions} minimized with the strategy of Section \ref{sec:LF_minimization_strategy}, including $C_{\lambda}$ from equation \eqref{eq:c_lambda}. The colored rows correspond to Figures \ref{fig:money_plot} for reference, where $\lambda = 0.3$.
\begin{table*}
\small 
    \centering 
\caption{Characterization of \texttt{SXS} simulations with \teob using different LFs, along with standardized $\egw$ measurements. Blue and red entries correspond to Figure \ref{fig:money_plot}. The leftmost column indicates the chosen LF, and the "Values" are computed from the \teob waveform that minimizes this LF. From left to right, the \texttt{SXS:BBH:XXXX} IDs are sorted by increasing $\egw$ from Table \ref{table:SXSdata}. Reference frequencies $f_0$ in Hz and $f_m=18$Hz correspond to total mass $M = m_1 + m_2 = 60 M_{\odot}$ and luminosity distance $D_L = 400$Mpc. Cases where the best-fit model has $\delta \egw \sim 100\%$ correspond to a preferred quasi-circular model by that LF, which occurs when $LF \sim \mm$ for the small eccentricity NR simulations.}
\definecolor{navyblue}{rgb}{0.0, 0.0, 0.5}
\def\nmet{4}

\begin{tabular}{ ||c|c| } 
\hline
\textbf{LF}& \textbf{Values}  \\
\hline

\multirow{\nmet}{*}{$\mm_{W}$} 
& $\mismatch_W$ \\ 
& $e_{0}$ \\ 
& $f_{0}$  \\ 
& $\degw(f_m)$ \\ 
\hline

\multirow{\nmet}{*}{$\mm_{L}$} 
& $\mathcal{M}_L$  \\ 
& $e_{0}$  \\ 
& $f_{0}$  \\ 
& $\degw(f_m)$  \\
\hline

\multirow{\nmet}{*}{$\hres$} 
& $e_{0}$  \\ 
& $f_{0}$  \\
& $\degw(f_m)$ \\ 
& $\mismatch_{W}$ \\
\hline

\multirow{\nmet}{*}{$|\Delta \egw (f_m)|$} 
&  $\degw(f_m)$ \\
& $e_{0}$\\ 
& $f_{0}$ \\ 
& $\mismatch_{W}$ \\
\hline

\multirow{\nmet}{*}{$\langle |\Delta \egw |\rangle$} 
& $e_{0}$  \\ 
& $f_{0}$  \\
& $\degw(f_m)$ \\ 
& $\mismatch_{W}$ \\
\hline

\multirow{\nmet}{*}{$C_{\lambda}$} 
& $e_{0}$  \\ 
& $f_{0}$  \\
& $\degw(f_m)$ \\ 
& $\mismatch_{W}$ \\
\hline
\end{tabular}%
\begin{tabular}{|*{10}{>{\rowfonttype}c|}}
\hline
 2294     & 2267     & 2270     & 2275     & 2280     & 2285     & 2290     & 2311     & 2300     & 2305     \\
\hline
\rowfont{\color{Red}}
 0.22\%    & 0.24\%    & 0.45\%    & 0.43\%    & 0.67\%    & 0.48\%    & 0.34\%    & 0.54\%    & 0.74\%    & 0.77\%    \\
 \rowfont{\color{Black}}
 0.000    & 0.000    & 0.027    & 0.073    & 0.103    & 0.130    & 0.162    & 0.226    & 0.218    & 0.262    \\
 14.203   & 13.958   & 14.070   & 14.041   & 14.280   & 14.921   & 16.098   & 14.514   & 14.936   & 15.201   \\
 \rowfont{\color{navyblue}}
 99.89\%   & 99.97\%   & 21.23\%   & 4.56\%    & 2.83\%    & 1.52\%    & 0.22\%    & 1.52\%    & 0.59\%    & 1.16\%    \\
 \hline
 \rowfont{\color{Black}}
 0.62\%    & 0.66\%    & 0.92\%    & 0.81\%    & 0.96\%    & 0.86\%    & 0.81\%    & 0.92\%    & 1.02\%    & 1.12\%    \\
 0.000    & 0.029    & 0.041    & 0.072    & 0.096    & 0.133    & 0.159    & 0.225    & 0.216    & 0.260    \\
 14.203   & 14.032   & 14.070   & 13.690   & 14.259   & 14.128   & 16.083   & 14.511   & 14.925   & 15.188   \\
 99.89\%   & 108.32\%  & 21.51\%   & 7.99\%    & 9.48\%    & 1.75\%    & 1.59\%    & 0.98\%    & 0.50\%    & 0.31\%    \\
 \hline
 0.012    & 0.011    & 0.000    & 0.035    & 0.083    & 0.109    & 0.145    & 0.228    & 0.218    & 0.253    \\
 13.690   & 13.352   & 12.400   & 13.044   & 13.494   & 14.481   & 15.537   & 13.802   & 14.146   & 15.136   \\
 240.90\%  & 20.54\%   & 100.00\%  & 57.13\%   & 25.45\%   & 16.51\%   & 13.01\%   & 2.78\%    & 4.48\%    & 2.16\%    \\
 1.61\%    & 1.25\%    & 0.94\%    & 2.36\%    & 2.77\%    & 3.83\%    & 5.02\%    & 3.67\%    & 4.04\%    & 2.56\%    \\
 \hline
 3.0E-07 & 1.5E-07 & 1.1E-07 & 1.3E-07 & 1.1E-07 & 2.3E-07 & 8.4E-07 & 3.2E-07 & 7.2E-07 & 2.3E-06 \\
 0.003    & 0.015    & 0.034    & 0.080    & 0.116    & 0.118    & 0.181    & 0.222    & 0.214    & 0.249    \\
 13.670   & 13.265   & 14.107   & 13.343   & 13.135   & 16.087   & 14.535   & 14.528   & 15.105   & 15.711   \\
 1.59\%    & 1.32\%    & 0.50\%    & 1.20\%    & 7.76\%    & 10.91\%   & 20.54\%   & 3.80\%    & 26.39\%   & 23.70\%   \\
 \hline
 0.002    & 0.015    & 0.031    & 0.084    & 0.102    & 0.119    & 0.154    & 0.258    & 0.203    & 0.265    \\
 14.666   & 13.101   & 15.231   & 12.836   & 14.755   & 15.994   & 16.836   & 12.699   & 15.839   & 14.901   \\
 0.12\%    & 0.09\%    & 0.08\%    & 0.01\%    & 0.00\%    & 0.02\%    & 0.00\%    & 0.01\%    & 0.00\%    & 0.01\%    \\
 1.45\%    & 1.04\%    & 3.30\%    & 3.26\%    & 3.22\%    & 7.42\%    & 14.13\%   & 13.01\%   & 4.47\%    & 25.15\%   \\
 \hline
 0.001    & 0.014    & 0.034    & 0.076    & 0.106    & 0.128    & 0.162    & 0.225    & 0.217    & 0.261    \\
 14.271   & 14.024   & 14.079   & 14.047   & 14.288   & 14.912   & 16.094   & 14.509   & 14.930   & 15.195   \\
 \rowfont{\color{navyblue}}
 8.81\%    & 0.61\%    & 0.20\%    & 0.04\%    & 0.01\%    & 0.06\%    & 0.06\%    & 1.05\%    & 0.29\%    & 0.94\%    \\
 \rowfont{\color{Red}}
 0.22\%    & 0.28\%    & 0.46\%    & 0.45\%    & 0.69\%    & 0.52\%    & 0.34\%    & 0.60\%    & 0.76\%    & 0.82\%    \\
\hline
\end{tabular}
    \label{table:summary}
\end{table*}

\begin{figure*}
\includegraphics[width=0.49\textwidth]{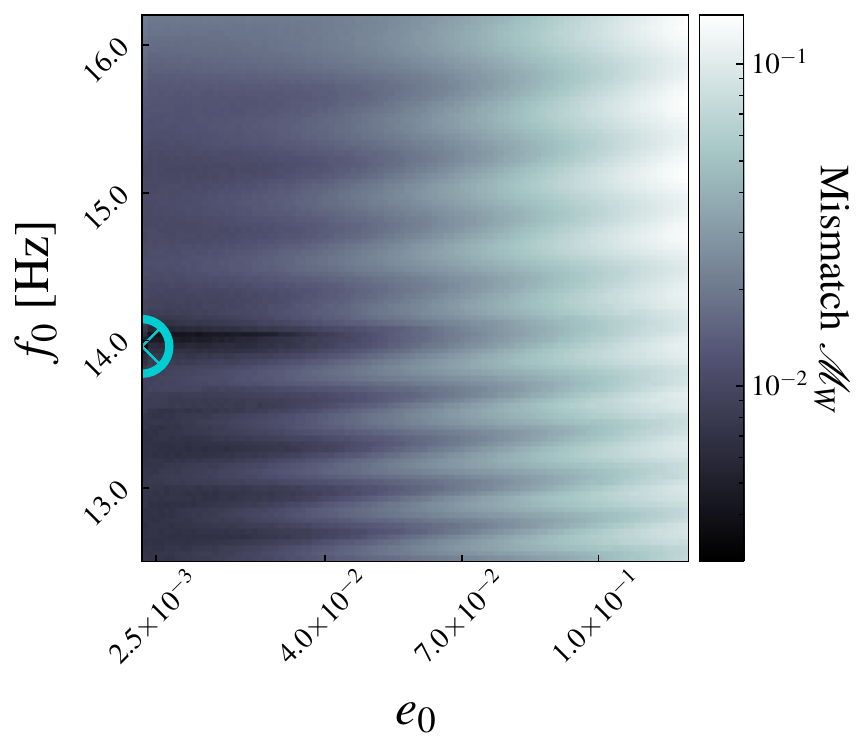}
\includegraphics[width=0.49\textwidth]{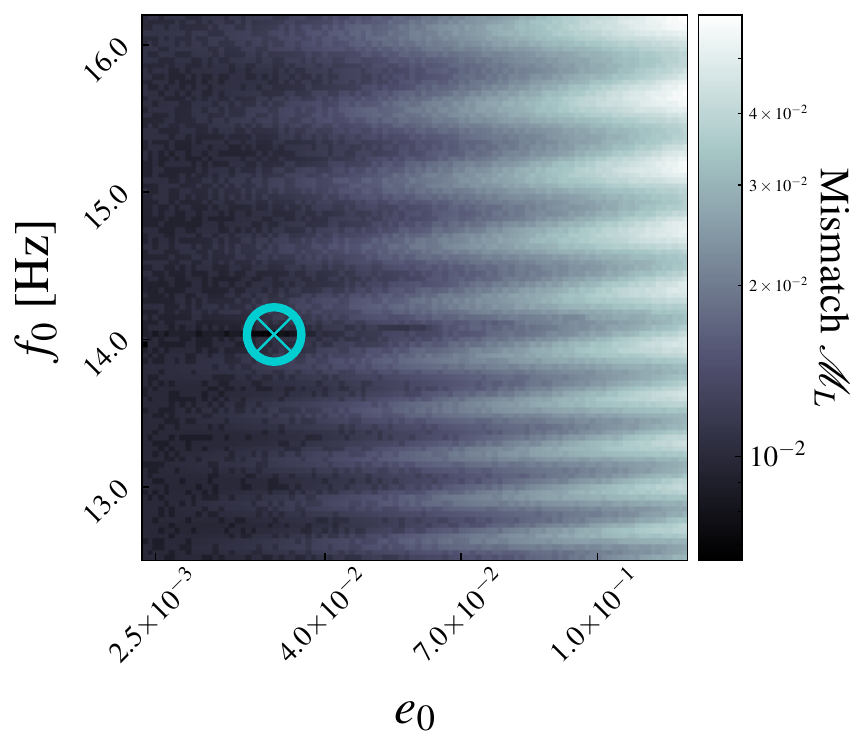}
    \caption{Mismatch based LFs with \texttt{SXS:BBH:2267} as the NR simulation, for two different PSDs $S_n$ (equation \ref{eq:overlap}). \underline{Left panel:} Mismatch $\mm_W$ with White Noise PSD.
    \underline{Right panel:} Mismatch $\mm_L$ With aLIGOZeroDetHighPower analytical PSD model from \cite{pycbc}. Physical quantities correspond to total mass $M = m_1 + m_2 = 60 M_{\odot}$ and luminosity distance $D_L = 400$Mpc.}.\label{fig:teob_mismatch_SXS002_2D}
\end{figure*}

For illustration, we also show the White Noise PSD mismatch $\mm_W$ and aLIGO PSD $\mm_L$ mismatch for \texttt{SXS:BBH:2270}, on the $(e_0,f_0)$ grid, in Figure \ref{fig:teob_mismatch_SXS002_2D}. In the White Noise case (left panel), \teob would actually prefer a quasi-circular computation $e_0=0$ to minimize the mismatch. The minimum would be different with the aLIGO model, which stresses the importance of choosing a representative PSD if one solely minimizes the mismatch from equation \ref{eq:mismatch} to characterize an eccentric simulation. Note that a more representative PSD to target observational applications rather than theoretical does not guarantee to solve model limitations, as the $\mm_L$ LF also favored a quasi-circular \teob waveform in one case (\texttt{SXS:BBH:2294}, in Table \ref{table:summary}).

\end{document}